\documentclass[prb, aps, 10pt, twocolumn, floatfix, superscriptaddress]{revtex4-1}

\usepackage{amssymb,amsfonts,amsmath,amsthm}
\usepackage{graphicx}
\usepackage[english]{babel}
\usepackage[latin1]{inputenc}
\usepackage[pdftex,colorlinks=true,pdfstartview=FitH,linkcolor=blue,citecolor=blue,urlcolor=blue]{hyperref}

\usepackage{bm}
\usepackage{float}
\usepackage{mathtools}
\usepackage{color,soul}
\usepackage{times}
\usepackage{upgreek}
\usepackage{MnSymbol}
\usepackage{verbatim}
\usepackage[bottom]{footmisc}
\usepackage{bbold}

\usepackage[dvipsnames,svgnames,table]{xcolor}
\usepackage{hyperref}
\usepackage[english]{babel}
\usepackage[caption=false]{subfig}

\hypersetup{
pdfstartview={FitH},
colorlinks=true,    
linkcolor=NavyBlue, 
citecolor=Blue,   
filecolor=NavyBlue, 
urlcolor=NavyBlue   
}

\newcommand{\bea}{\begin{eqnarray}}
\newcommand{\eea}{\end{eqnarray}}
\newcommand{\be}{\begin{equation}}
\newcommand{\ee}{\end{equation}}

\newcommand{\xv}{\hat{\bm{x}}}
\newcommand{\yv}{\hat{\bm{y}}}
\newcommand{\zv}{\hat{\bm{z}}}
\newcommand{\dt}[1]{\frac{d #1}{dt} }

\usepackage{cancel}

\begin{document}

\title{Solid-state continuous time crystal with a built-in clock}

\author{I. Carraro Haddad}
\thanks{These two authors contributed equally}
\affiliation{Centro At{\'{o}}mico Bariloche and Instituto Balseiro,
Comisi\'on Nacional de Energ\'{\i}a At\'omica (CNEA)- Universidad Nacional de Cuyo (UNCUYO), 8400 Bariloche, Argentina.}
\affiliation{Instituto de Nanociencia y Nanotecnolog\'{i}a (INN-Bariloche), Consejo Nacional de Investigaciones Cient\'{\i}ficas y T\'ecnicas (CONICET), Argentina.}

\author{D. L. Chafatinos}
\thanks{These two authors contributed equally}
\affiliation{Centro At{\'{o}}mico Bariloche and Instituto Balseiro,
Comisi\'on Nacional de Energ\'{\i}a At\'omica (CNEA)- Universidad Nacional de Cuyo (UNCUYO), 8400 Bariloche, Argentina.}
\affiliation{Instituto de Nanociencia y Nanotecnolog\'{i}a (INN-Bariloche), Consejo Nacional de Investigaciones Cient\'{\i}ficas y T\'ecnicas (CONICET), Argentina.}

\author{A.~S. Kuznetsov}
\affiliation{Paul-Drude-Institut f\"{u}r Festk\"{o}rperelektronik, Leibniz-Institut im Forschungsverbund Berlin e.V., Hausvogteiplatz 5-7,\\ 10117 Berlin, Germany.}

\author{I. Papuccio}
\affiliation{Centro At{\'{o}}mico Bariloche and Instituto Balseiro,
Comisi\'on Nacional de Energ\'{\i}a At\'omica (CNEA)- Universidad Nacional de Cuyo (UNCUYO), 8400 Bariloche, Argentina.}
\affiliation{Instituto de Nanociencia y Nanotecnolog\'{i}a (INN-Bariloche), Consejo Nacional de Investigaciones Cient\'{\i}ficas y T\'ecnicas (CONICET), Argentina.}

\author{A.~A. Reynoso}
\affiliation{Centro At{\'{o}}mico Bariloche and Instituto Balseiro,
Comisi\'on Nacional de Energ\'{\i}a At\'omica (CNEA)- Universidad Nacional de Cuyo (UNCUYO), 8400 Bariloche, Argentina.}
\affiliation{Instituto de Nanociencia y Nanotecnolog\'{i}a (INN-Bariloche), Consejo Nacional de Investigaciones Cient\'{\i}ficas y T\'ecnicas (CONICET), Argentina.}

\author{A. Bruchhausen}
\affiliation{Centro At{\'{o}}mico Bariloche and Instituto Balseiro,
Comisi\'on Nacional de Energ\'{\i}a At\'omica (CNEA)- Universidad Nacional de Cuyo (UNCUYO), 8400 Bariloche, Argentina.}
\affiliation{Instituto de Nanociencia y Nanotecnolog\'{i}a (INN-Bariloche), Consejo Nacional de Investigaciones Cient\'{\i}ficas y T\'ecnicas (CONICET), Argentina.}

\author{K. Biermann}
\affiliation{Paul-Drude-Institut f\"{u}r Festk\"{o}rperelektronik, Leibniz-Institut im Forschungsverbund Berlin e.V., Hausvogteiplatz 5-7,\\ 10117 Berlin, Germany.}

\author{P.~V. Santos}
\email[Corresponding author, e-mail: ]{santos@pdi-berlin.de}
\affiliation{Paul-Drude-Institut f\"{u}r Festk\"{o}rperelektronik, Leibniz-Institut im Forschungsverbund Berlin e.V., Hausvogteiplatz 5-7,\\ 10117 Berlin, Germany.}

\author{G. Usaj}
\email[Corresponding author, e-mail: ]{usaj@cab.cnea.gov.ar}
\affiliation{Centro At{\'{o}}mico Bariloche and Instituto Balseiro,
Comisi\'on Nacional de Energ\'{\i}a At\'omica (CNEA)- Universidad Nacional de Cuyo (UNCUYO), 8400 Bariloche, Argentina.}
\affiliation{Instituto de Nanociencia y Nanotecnolog\'{i}a (INN-Bariloche), Consejo Nacional de Investigaciones Cient\'{\i}ficas y T\'ecnicas (CONICET), Argentina.}
\affiliation{TQC, Universiteit Antwerpen, Universiteitsplein 1, B-2610 Antwerpen, Belgium}
\affiliation{CENOLI, Universit\'e Libre de Bruxelles - CP 231, Campus Plaine, B-1050 Brussels, Belgium}

\author{A. Fainstein}
\email[Corresponding author, e-mail: ]{afains@cab.cnea.gov.ar}
\affiliation{Centro At{\'{o}}mico Bariloche and Instituto Balseiro,
Comisi\'on Nacional de Energ\'{\i}a At\'omica (CNEA)- Universidad Nacional de Cuyo (UNCUYO), 8400 Bariloche, Argentina.}
\affiliation{Instituto de Nanociencia y Nanotecnolog\'{i}a (INN-Bariloche), Consejo Nacional de Investigaciones Cient\'{\i}ficas y T\'ecnicas (CONICET), Argentina.}

\date{\small \today}
\begin{abstract}
{Time crystals (TCs) are many-body systems displaying spontaneous breaking of time translation symmetry. Here, we demonstrate a TC using driven-dissipative condensates of microcavity exciton-polaritons, spontaneously formed from an incoherent particle bath. In contrast to other realizations, the TC phases can be controlled by the power of continuous-wave non-resonant optical drive exciting the condensate and optomechanical interactions with phonons. Those phases are for increasing power: (i) Larmor precession of pseudo-spins - a signature of continuous TC, (ii) locking of the frequency of precession to self-sustained coherent phonons - stabilized TC, (iii) doubling of TC frequency by phonons - a discrete TC with continuous excitation. These results establish microcavity polaritons as a platform for the investigation of time-broken symmetry in non-hermitian systems.
}
\end{abstract}

\maketitle


Symmetries govern the behavior of a physical system. Ordinary material crystals constitute, however, a well-known example in which a quantum many-body system spontaneously breaks space translation symmetry. Similarly, time crystals have been proposed as quantum systems exhibiting spontaneous breaking of the time translation symmetry (TTS)~\cite{Wilczek2012}. It was soon realized, however, that time-independent Hamiltonians necessarily lead to stationary ground states~\cite{Bruno2013}. This conclusion, far from closing the subject, stimulated the search for conditions under which time-crystals could still emerge. One direction explores periodically-driven many-body Hamiltonians~\cite{Sacha2018,Else2020},  leading to stable states with periods distinct from the drive, often identified by period doubling (PD). These ``discrete time crystals'' (DTCs) have been observed in diverse physical systems including cold atoms~\cite{Kessler2021}, magnons in superfluid $^3$He~\cite{Autti2018,Autti2021}, nuclear spins~\cite{Zhang2017}, and photonic devices~\cite{Taheri2022}. Searches for a  time crystal closer to the original idea of a time-dependent ground state of a many-body quantum system in the absence of time-dependent driving, have however not waned.

\begin{figure*}[!hht]
 \begin{center}
    \includegraphics[trim = 0mm 0mm 0mm 0mm,clip=true, keepaspectratio=true, width=1.4 \columnwidth,angle=0]{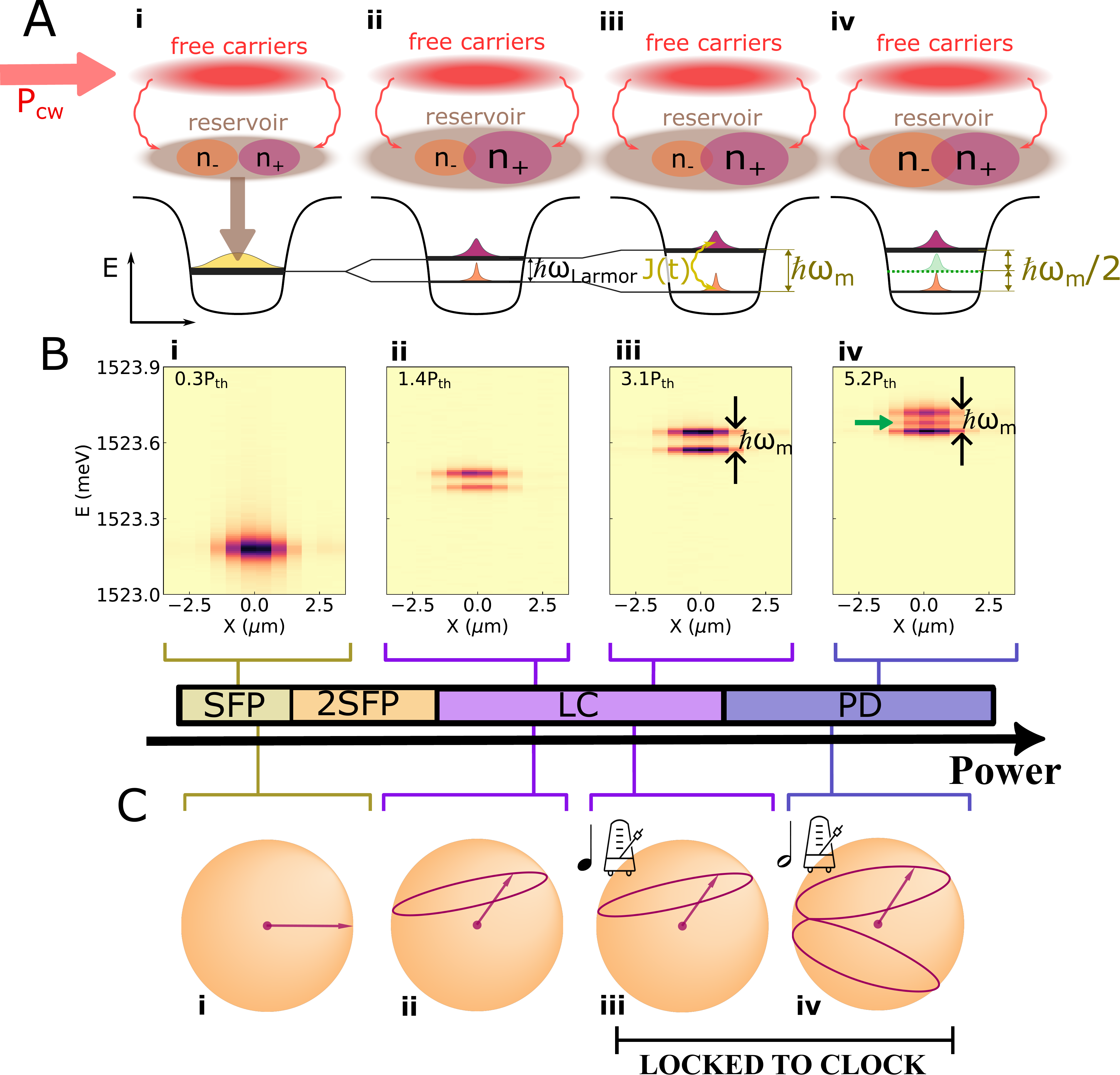}
\end{center}
\vspace{-0.8 cm}
\caption{\textbf{Spinor Time Crystal with Internal Clock.} \textbf{(A)} Experimental setup and dynamics with increasing non-resonant cw laser power $P_\mathrm{cw}$. \textbf{(B)} Spectrally-resolved photoluminescence images of the $3 \times 3~\mu$m$^2$ square trap for the four powers in \textbf{(A)}.
A continuous wave linearly polarized pump laser non-resonantly injects free carriers with $P\mathrm{cw}$. These relax, forming an exciton reservoir feeding a polariton trap. Due to linear-polarized pumping and carrier relaxation, at low powers, exciton spin-up and spin-down populations in the reservoir are balanced, leading to a degenerate polariton ground state (\textbf{(A-B)i}). Increasing $P_\mathrm{cw}$ breaks time translation symmetry, initiating spontaneous Larmor precession (continuous time crystal), indicated by pseudospin state splitting (\textbf{(A-B)ii}). Polariton dynamics induces coherent mechanical oscillation, back-acting to lock Larmor precession to confined phonon frequency $\omega_m/2\pi \sim 20$GHz (\textbf{(A-B)iii}). Further power increase results in self-induced mechanical oscillation leading to period doubling (sidebands split by $\omega_m/2$) (\textbf{(A-B)iv}).
\textbf{(C)} Spinor representation for states evolving from single or double stable fixed points (\textbf{i}, SFP and 2SFPs, respectively), to a limit cycle (LC) corresponding to Larmor precession with either a self-determined (\textbf{ii}) or locked (\textbf{iii}) frequency, and finally to period doubling (PD) locked to the mechanical clock (\textbf{iv}). 
}
\label{Fig1}
\end{figure*}

It has been theoretically argued that TTS can naturally break in an open quantum system where dissipation stabilizes time-crystal dynamics compensating for losses through continuous pumping.~ \cite{Kessler2021,Autti2018,Gong2018,Iemini2018,Tucker2018,Zhu2019,Buca2019,Lazarides2020} 
These systems are termed ``continuous time crystals'' (CTCs), as their Hamiltonians are time-independent.  Probably the closest experimental implementation of such a CTC is the recent report of an open atom-cavity system that oscillates between two checkerboard density patterns, when destabilized by an additional quasi-resonant continuous laser drive~ \cite{Kessler2019,Kongkhambut2022}. 
We report here a fully CTC implemented in a driven-dissipative exciton-polariton system.~\cite{Nalitov2019}  Contrary to all previous proposals, this solid-state many-body quantum time crystal spontaneously forms from an incoherent particle bath that is loaded with a continuous non-resonant optical drive. Intriguingly, we demonstrate period doubling concerning an internal mechanical clock in this CTC state.

 \paragraph*{\textbf{A polaromechanical continuous time crystal loaded from an incoherent reservoir}} 

Our solid-state CTC system is based on polaritons, quasiparticles resulting from the strong coupling between excitons and photons in a semiconductor intracavity trap. These are bosonic excitations that display a plethora of notable phenomena which derive from their mutual Coulomb interactions and dissipative nature (due to the excitonic and photonic component, respectively).~\cite{Hartmann2006,Amo2009,CarusottoRMP2013} 
Being composite bosons, they transition to a non-equilibrium Bose-Einstein condensate under appropriate conditions~\cite{Kasprzak2006}. Our traps also confine $\sim 20$~GHz mechanical vibrations that very efficiently couple to polaritons through deformation potential interaction~\cite{Fainstein2013,Santos2023}. In a recent experiment we showed that two polariton traps with energy levels detuned at the phonon energy lead to self-oscillation of a coherent mechanical wave (phonon lasing)~\cite{Chafatinos2020,Reynoso2022}. Here, we reveal that the polariton ground state in one such trap can develop a non-linear self-sustained dynamics, intimately affected by mechanics in ways that expose characteristics of both continuous and discrete time crystals.

As depicted in Fig.~\ref{Fig1}(A), a semiconductor solid-state device designed to trap polaritons is subject to non-resonant continuous-wave (cw) optical excitation. Electron-hole particles are optically injected at high energies, and left to relax and find their stationary dynamics. Energy loss occurs through emission of phonons, erasing phase information set by the drive until a relaxation bottleneck leads to exciton reservoir build-up (depicted at the top of traps in Fig.\ref{Fig1}(A)). This reservoir is, a priori, unpolarized by construction with equal population of spin-up $n_{+}$ and spin-down $n_{-}$ excitons generated by the non-resonant linearly polarized laser. This reservoir acts as the bath feeding the polariton ground state (GS) of a nominally square $3 \times 3~\mu$m$^2$ trap fabricated by laterally micropatterning the spacer region of an otherwise planar microcavity~\cite{Winkler2015,Kuznetsov2018} (see S1 of the supplementary material~\cite{SM}) .

\begin{figure*}[ht]
    \centering 
\includegraphics[trim = 0mm 0mm 0mm 0mm, clip=true, keepaspectratio=true, width=1.8\columnwidth,angle=0]{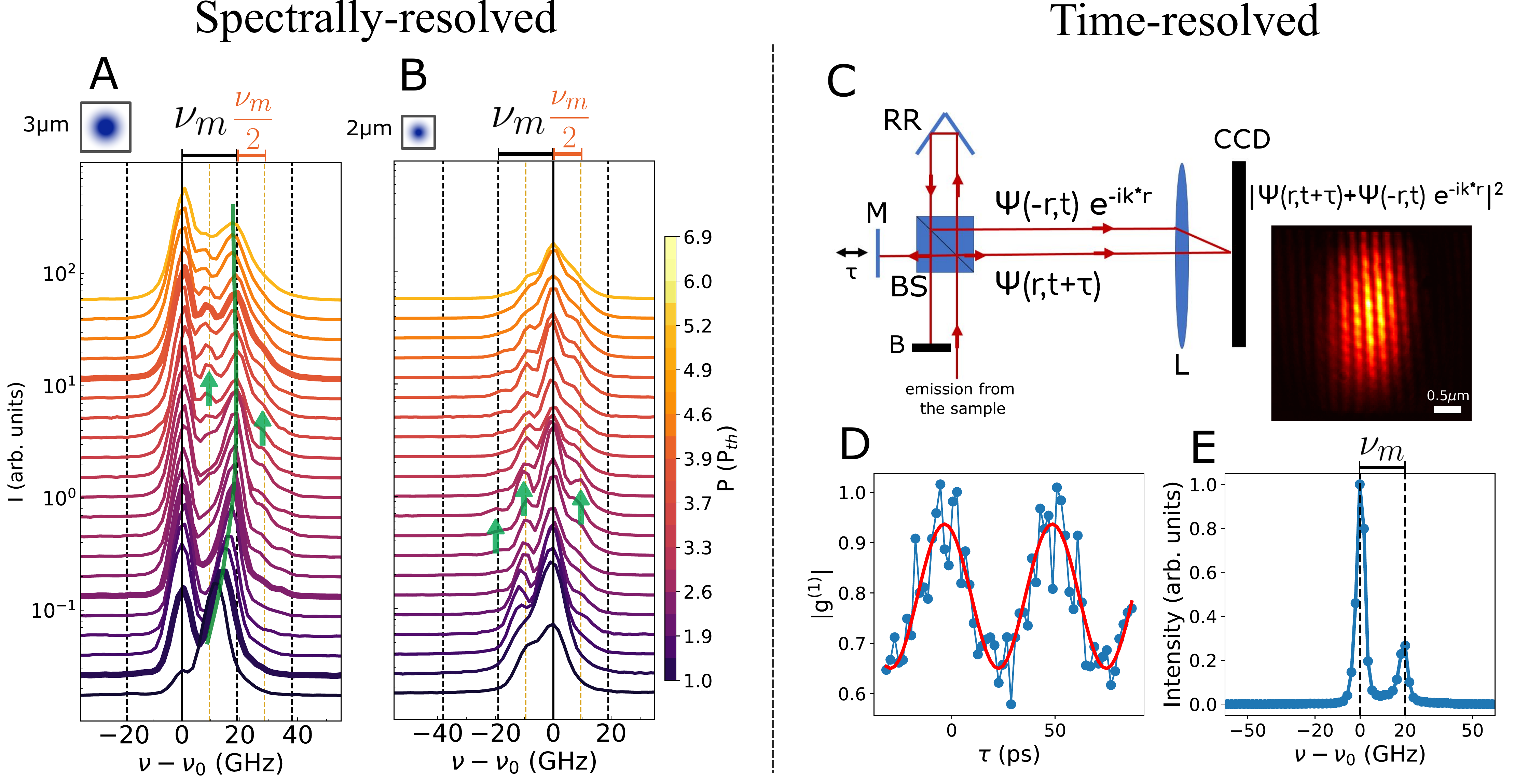}
 \caption{
\textbf{Spectra and Time-Resolved Coherence Function $\mathbf{g^{(1)}(r,\tau)}$.}
\textbf{(A)} and \textbf{(B)} Spectra examples for increasing continuous wave excitation power (bottom to top) in individual $3 \times 3~\mu$m$^2$ and $2 \times 2~\mu$m$^2$ square traps, respectively. Thicker lines in \textbf{(A)} highlight spectra corresponding to spatial images in Fig.\ref{Fig1}(b). Frequencies are referenced to the most intense peak ($\nu_0$), subtracting the interaction-induced blue-shift. Vertical dashed lines indicate multiples of the phonon frequency ($\nu_\mathrm{m} \sim 20$GHz) and its half ($\nu_\mathrm{m}/2$). Up-arrows mark peaks related to period-doubling. \textbf{(C)} Experimental set-up scheme for off-axis digital holography and image obtained for a given $\tau$ defined by the mirror M position. Light collected from the polariton condensate ground state is split by a non-polarizing beamsplitter BS, with $50\%$ directed to a retroreflector RR and the other $50\%$ to mirror M on a delay line. Reflected light back to the sample is blocked in B, creating an interference pattern on a liquid-$N_2$ cooled CCD through lens L. \textbf{(D)} Example of time-dependent $g^{(1)}(r,\tau)$ for a $2 \times 2\mu$m$^2$ square trap with Larmor precession locked to the phonon frequency $\omega_m/2\pi \sim$20GHz, along with corresponding spectra in \textbf{(E)}.
    }
\label{Fig2}
\end{figure*} 

At low continuous excitation powers $P_\mathrm{cw}$ a single confined level of the trap is observed [cf Figs.\ref{Fig1}(A-B)i]. On increasing $P_\mathrm{cw}$ a condensation occurs at a threshold $P_\mathrm{th}$, signaled by a strong non-linear increase in the level occupation, accompanied by a narrowing of the line down to a fraction of a GHz (coherence times of a few nanoseconds) as well as by an energy blue-shift determined by polariton-polariton and polariton-reservoir exciton-mediated interactions. On further increasing $P_\mathrm{cw}$ a series of striking qualitative changes occur that are illustrated in Figs.~\ref{Fig1}(A) and (B).  These are: (panels ii) the spontaneous splitting of lines representing a two-level system. These lines are attributed to the $\sigma=\pm$ polarization degree of freedom of the condensate, which is directly linked to the spin of the excitons and to the polarization of the cavity photons through angular momentum selection rules.~\cite{Shelykh2005,Ohadi2015,Gnusov2020,Siggurdsson2020} (iii) The subsequent locking of the splitting to the frequency of the acoustic phonon confined in the same trap $\nu_m = \omega_m/2\pi \sim 20$GHz; (iv) and, finally, the emergence of sidebands separated by $\nu_m/2 \sim 10$GHz signaling a period doubling respect to $\nu_m$.  Interestingly,  this two-level system can be interpreted by the dynamics of a spinor in the presence of synthetic fields, as illustrated in Fig.\ref{Fig1}(C). These orbits, which reflect self-induced time crystalline oscillations of the condensate  polarization, will be further explained with the theory below.

 \paragraph*{\textbf{Experimental evidence of spontaneous breaking of time translation symmetry}} 
 
The sequence of spectra leading to the spatial images in Fig.\ref{Fig1}(B) are shown in Fig.\ref{Fig2}(A) for increasing cw excitation power $P > P_\mathrm{th}$ (from bottom to top).  Another example, obtained for a $2 \times 2~\mu$m$^2$ square trap, is shown in Fig.\ref{Fig2}(B). Both structures display a slightly different behavior, but the phenomenology is broadly the same. Above threshold the mode is observed to split in two, with the splitting increasing with power. When $P \sim 2P_\mathrm{th}$ the mode separation locks to the phonon frequency (Fig.~\ref{Fig2}(A)) or to half of it (Fig.~\ref{Fig2}(B)), and after this locking sidebands emerge separated by $\nu_\mathrm{m}/2$.  These sidebands, reflecting PD referred to the phonon period, are indicated with up-arrows in the figures (see S7 of the supplementary material for additional experimental examples of CTC behavior~\cite{SM}).

The observed sidebands are definitive ``smoking-gun'' evidence for the emergence of a coherent time dynamics involving and modulating the exciton-polaritons~\cite{Rayanov2015}, that is, of a time crystal. Additionally, we obtain direct information of the time-dependence by measuring the time-resolved spatial first-order coherence function $g^{(1)}(r,\tau)$ as schematized in Fig.~\ref{Fig2}(C) [see S1 of the supplementary material~\cite{SM}]. $g^{(1)}(r,\tau)$, illustrated for a situation close to  Fig.~\ref{Fig1}(B)III, is presented  in Fig.~\ref{Fig2}(D). It displays clear temporal oscillations with a period that exactly matches the inverse of the phonon frequency $1/\nu_\mathrm{m} \sim 50$~ps, and is consistent with the spectral line-splitting as simultaneously obtained and presented in Fig.~\ref{Fig2}(E).

\begin{figure*}[ht]
    \centering 
    \includegraphics[trim = 0mm 0mm 0mm 0mm,clip=true, keepaspectratio=true, width=2\columnwidth,angle=0]{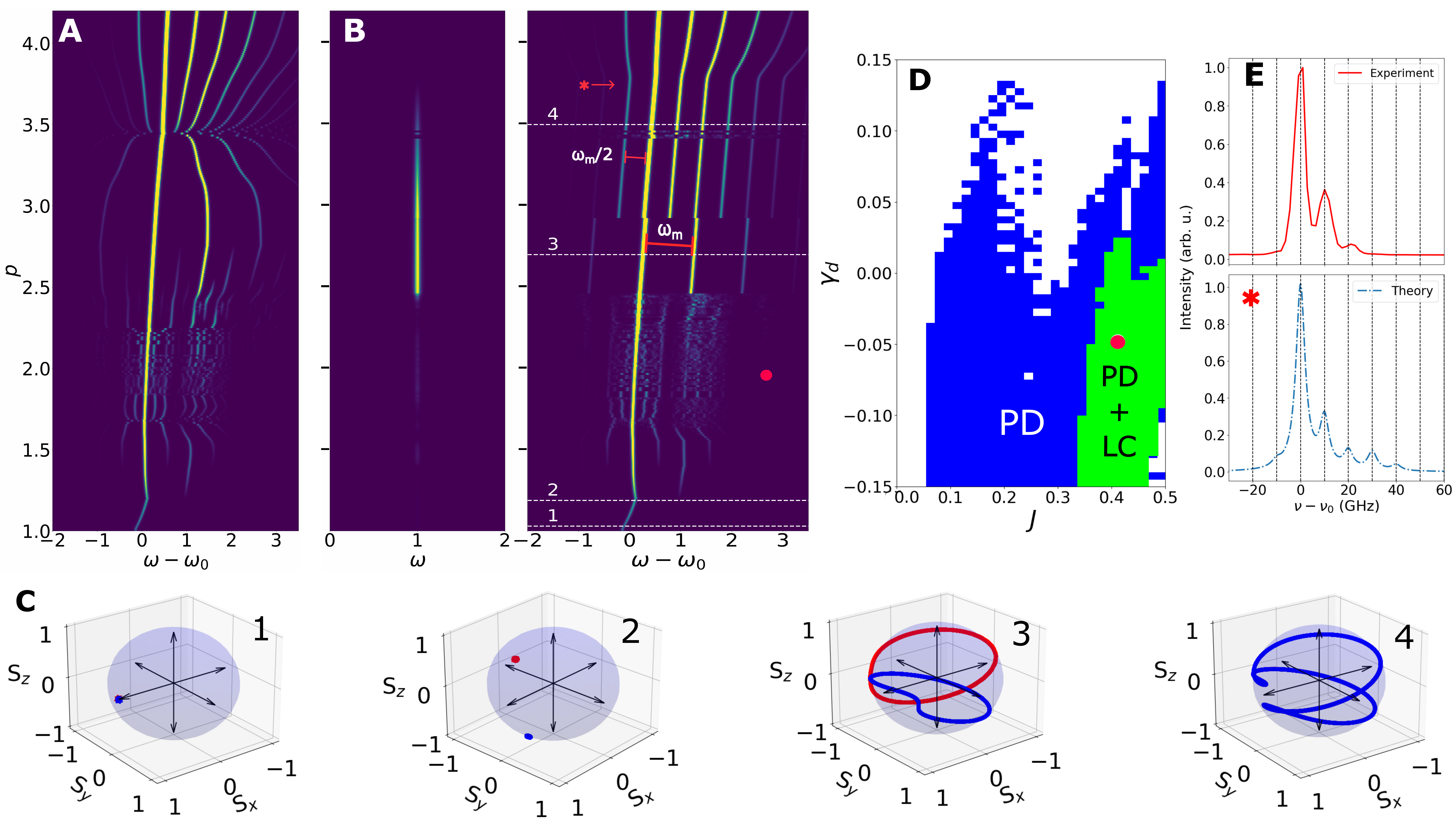}
    \caption{
    \textbf{Theory of Continuous Polaromechanical Time Crystal.} \textbf{(A)} and \textbf{(B)} show energy spectrum color maps from Eq. \eqref{eq: S} as a function of $p=P_\mathrm{cw}/P_\mathrm{th}$ without and with applied mechanics, respectively ($P_\mathrm{th}$ is the condensation threshold power). Simulation parameters: $2J=0.45$, $\Delta\varepsilon=0$, $\hbar \gamma = 0.85$, $2\hbar \gamma_d = -0.05$, $U_0=0.25$, $U_0^R = 3.6$, $\hbar \gamma_R = 0.4$, and $g_0=3.5 \times 10^{-5}$ (scales in terms of $\omega_\mathrm{m} = 2\pi 20$~GHz). In \textbf{(B)}, the left panel shows phonon spectra, while the right panel displays polariton spectra around $\hbar \omega_0$. Horizontal dashed white lines in \textbf{(B)} indicate powers corresponding to orbits on normalized Poincare spheres in \textbf{(C)}. Red and blue denote distinct initial conditions, highlighting dynamical attractors. \textbf{(D)} Dynamical phase diagram as a function of real ($J$) and imaginary ($\gamma_d$) polariton coupling components. Green or blue indicates regions where, as a function of $p$, limit cycles locked at $\omega_\mathrm{m}/2\pi = 20$~GHz (LC) and period doubling (PD) or only PD are observed. Red dot indicates parameters used in \textbf{(B)}. \textbf{(E)} Example of calculated (bottom) and experimental (top) spectra, the former for the power indicated with an asterisk in \textbf{(B)}.
}
    \label{Fig3}
\end{figure*} 

 \paragraph*{\textbf{Theoretical model of a polariton condensate coupled to a dynamical reservoir and a mechanical clock}} 
 
The results in Figs.\ref{Fig1} and \ref{Fig2} emerge from the intricate interplay of non-resonant driving, coupling to excitons in a dynamical reservoir, cavity dissipation, interactions in a many-body quantum polariton system, and the tailored coupling and feedback with confined phonons. In the condensation regime the occupation is large, and number fluctuations can be in a first approximation neglected. This leads to a semiclassical description in terms of a coherent state, and its coupling to a mean-field bath. To describe this system we use the generalized Gross-Pitaevskii equations (gGPEs) for the condensate field $\tilde{\psi}_\sigma$ and reservoir population $\tilde{n}_\sigma$:~\cite{CarusottoRMP2013,Ohadi2015,Gnusov2020,Siggurdsson2020,Rayanov2015,Wouters2008}
\begin{eqnarray}
\label{eq:renormalized_plus1}
\nonumber
i\hbar\dt{\tilde{\psi}_\sigma} &=&(\varepsilon_\sigma+U_0|\tilde{\psi}_\sigma|^2+U^R_0 \tilde{n}_\sigma)\tilde{\psi}_\sigma-(J+i\hbar\gamma_d)\tilde{\psi}_{\bar{\sigma}}\\
&&+\frac{i\hbar\gamma}{2}(\tilde{n}_\sigma-1)\tilde{\psi}_\sigma\,,\\
\label{eq:renormalized_plus2}
\nonumber
\dt{\tilde{n}_\sigma}&=&\gamma_R\left(p_\sigma-\left(1+|\tilde{\psi}_\sigma|^2\right)\tilde{n}_\sigma \right)\,.\\   
\end{eqnarray}
In these equations $\tilde{\psi}_\sigma$ and $\tilde{n}_\sigma$ have been scaled with the corresponding polariton and reservoir average populations $\sqrt{\rho_0}$ and $n_0$, respectively. $\varepsilon_\sigma$ is the polariton bare energy, and $U_0$ and $U_0^R$ account for the same-spin Coulomb interaction between polaritons and with excitons in the reservoir, respectively (see S2 of the supplementary material for more details on the gGPEs~\cite{SM}). The complex coupling $-(J+i\hbar\gamma_d)$ contains two contributions : (i) a Josephson-like coupling ($-J$), that arises from the energy splitting of the $X$ and $Y$ linearly polarized states, and (ii) a dissipative coupling  ($-i\hbar\gamma_d$)  originated on the difference between the effective decay rates of the $X$-$Y$ linear polarized modes~\cite{Aleiner2012} (see S8 of the supplementary material for experimental proof of different linewidth of the spinor modes~\cite{SM}). Here $\psi_X = (\tilde{\psi}_+ + \tilde{\psi}_-)/\sqrt{2}$ and $ \psi_Y= (\tilde{\psi}_+ - \tilde{\psi}_-)/\sqrt{2}$ represent the linear polarized base, natural due to the symmetry of the trap.  The last term in Eq.~\eqref{eq:renormalized_plus1} represents the stimulated gain from the exciton's reservoir (proportional to $\tilde{n}_\sigma$), and the bare polariton decay rate ($\gamma$). Eq.~\eqref{eq:renormalized_plus2} describes the dynamics of the reservoir, which is controlled by the pump power (here $p_\sigma=P_\sigma/P_\mathrm{th}$, with $P_\sigma$ the external cw driving of the $\sigma$ reservoir, and $P_\mathrm{th}=\gamma\gamma_R/R$ is the condensation threshold power defined by the exciton's non-radiative decay rate ($\gamma_R$) and the stimulated decay to the condensate $R$ (spin-flip between the reservoirs is not included). The coupling to the phonons can be introduced by assuming that either $\varepsilon_\sigma$, $J$, or both depend on the vibrational coordinate $x$, and that the latter is affected by the polariton dynamics through the optical forces,~\cite{Sesin2023} thus providing the bridge with the physics of cavity optomechanics~\cite{RMP}.

A clear physical picture of the resulting dynamics can be grasped by changing to a spinor description in the Poincare sphere~\cite{Chestnov2021}  via a pseudospin transformation $S_j= \tilde{\Psi}^\dagger \sigma_j \tilde{\Psi}$, where $\sigma_j$ are the different Pauli matrices. Each component of the pseudospin takes the form $S_x = 2 \mathrm{Re} \left( \tilde{\psi}_+\tilde{\psi}_-^* \right)=2 \cos(\theta)\sqrt{N_+N_-}$, $S_y = 2 \mathrm{Im} \left( \tilde{\psi}_+ \tilde{\psi}_-^* \right)=2 \sin(\theta)\sqrt{N_+N_-}$ and $S_z = \left| \tilde{\psi}_+ \right|^2 - \left| \tilde{\psi}_- \right|^2=N_+-N_-$. 
Here $N_\sigma$ and $\theta_\sigma$ are defined by the Madelung transformation, $\tilde{\psi}_\sigma = \sqrt{N_\sigma} e^{i \theta_\sigma}$. The polarization of the photon component of the polaritons maps directly with their pseudospin, with the magnitude and sign of $S_x$, $S_y$ and $S_z$ being the horizontal, diagonal or circular right polarization of the emitted light (vertical, anti-diagonal or circularly left, if the components are negative). Using this notation the following set of equations are obtained~\cite{Chestnov2021}:
\begin{equation}
    \dt{\bm{S}}=\frac{1}{\hbar}\bm{B}\times\bm{S}+S\bm{E}+ \Gamma\bm{S}\,,
        \label{eq: S}
\end{equation}
where
\begin{eqnarray}
\nonumber
\bm{B}&=&2J\,\xv-(U_0S_z+2U_0^Rm)\,\zv\,,\\
\nonumber
\bm{E}&=&-2\gamma_d\,\xv+\gamma m\,\zv\,,\\
\Gamma&=&\gamma(n-1)\,,
\label{eq: BE}
\end{eqnarray}
with $n=(\tilde{n}_++\tilde{n}_-)/2$ and $m=(\tilde{n}_+-\tilde{n}_-)/2$, and where we have assumed $\varepsilon_{+}=\varepsilon_{-}$, that is no external magnetic field is present.

Equation~\eqref{eq: S} shows that the condensate spinor follows dynamical equations resembling those in magnetic resonance phenomena, albeit with some important differences. The first term with the vector product leads to Larmor precession. Note that the mechanical modulation $J(t)$ can allow for the coherent control strategies based on pulsed resonant fields. The last term mimics the phenomenological way in which spin decay times are usually introduced, where $\Gamma$ describes the net gain of the system. The second term can be associated with a synthetic crystal-field $\bm{E}$ favoring specific spinor orientations. Equation~\eqref{eq: BE} shows that $\bm{B}$ has two components, one in the $x$ direction, originated in the Josephson coupling $J$ (i.e., in the energy splitting of the $X$-$Y$ modes), and another along $z$ which is non-linear and changes with the population imbalance of the $(+,-)$ modes ($S_z$) and that of the reservoirs ($m$). 
This implies that, in general, the Larmor precession can lead to a time dependence of $\bm{B}$. This feedback  is critical first to induce a spontaneous polarization of the condensate, and then for establishing a self-sustained dynamics.
The second line in Eq.~\eqref{eq: BE} shows that the field $\bm{E}$ has an $x$ component reflecting the coupling caused by the difference of dissipation between the $X$ and $Y$ modes, and a $z$ component deriving from the difference in gain between the $(+,-)$ modes caused by the reservoir imbalance $m$. It involves again a non-linearity, and thus can be time-dependent if a self-sustained dynamics sets in.  A coupling between the dynamics of the reservoir bath and that of the polaritons in the trap, is critical for observing the time crystal behavior in our system (see S3-S4 of the supplementary material for a more detailed analysis of the spinor model~\cite{SM}).

We account for the effect of the mechanical vibrations on the polaritons through a periodic modulation of the Josephson-like coupling $J(t) = J_0 + \hbar g_0 x(t)$, where $J_0$ is the constant coupling described before, $g_0$ is the linear optomechanical coupling constant and $x(t)$ is the phonons' displacement
self-consistently determined by solving for the mechanical dynamics coupled to the polariton equations Eqs.~\eqref{eq:renormalized_plus1}-\eqref{eq:renormalized_plus2}~\cite{Chafatinos2020,Reynoso2022}. If a threshold situation occurs in which 
a self-induced coherent mechanics sets-in, this leads to an oscillating $\bm{B}(t)$ field, with an $x$ component which approximately will follow: $B_x(t) = 2J_0 + 2J_p \cos{(\omega_\mathrm{m} t)}$, where $J_p = \hbar g_0 \sqrt{n_p}$. $n_p$ is the number of phonons, and $\omega_\mathrm{m}$ is the phonon frequency (see S5 of the supplementary material~\cite{SM}).

 \paragraph*{\textbf{Continuous time crystalline phases tuned by the particle number}} 

In Figs. \ref{Fig3}(A) and (B) we compare the dynamics without the phonons ($J_p=0$) and with them ($J_p \neq 0$), respectively. These calculations were obtained for $p_+=p_-=p$. The shown colormaps display the energy spectrum as a function of normalized laser power $p$. The left panel in (b) corresponds to the spectral region around the confined phonon frequency, while the right one maps the region corresponding to the polariton modes around the bare frequency $\omega_0$. Frequencies are scaled to the confined phonon frequency $\omega_\mathrm{m}/2\pi = 20$~GHz. The phonon spectra display a single line at $\omega_\mathrm{m}$, with an intensity that directly reflects the emergence of a self induced coherent mechanical oscillation (observed mainly around $2.4 \lesssim p \lesssim 3.7$). For the polariton spectra, besides a blue-shift of the spectra, mainly determined by the reservoir average population ($n$) through $U_0^R$, a rich scenario develops with increasing external pumping.  Concentrating first in Fig. \ref{Fig3}(B) and the corresponding Poincare spheres in Figs. \ref{Fig3}(C), the sequence goes from: (1) the formation of synchronized condensates at low pumping (one stable fixed point SFP)~\cite{Wouters2008}; (2) a symmetry-breaking pitchfork bifurcation~\cite{Hamel2015} to two SFPs of the same frequency, with unequal occupation of the condensates of the two pseudospins $\sigma = \pm 1$ (emergence of a synthetic magnetic field), identified by red and blue colors and which are selectively accessed depending on the initial conditions; (3) loss of the stability of the SFPs in the system resulting in two possible limit cycle (LC) dynamics (self-induced Larmor precession)~\cite{Sigurdsson2022} that lock to the phonon frequency $\omega_\mathrm{m}$ (point at which the Larmor precession drives the mechanics),~\cite{Chafatinos2020,Reynoso2022}; and (4) a transition to a PD dynamics again locked at half the phonon frequency $\omega_\mathrm{m}/2$. The latter arises from the merging of the two possible orbits in (3).

On further increasing the excitation power the system may unlock its Larmor frequency from the mechanical ``clock'' (illustrated above $p \sim 3.7$ in Fig.~\ref{Fig3}(B)), and even at higher powers eventually it can attain again a synchronized state (SFP) via a Hopf bifurcation from the limit cycle with polarization orthogonal to that in (1) (see S6 of the supplementary material~\cite{SM}). We note that depending on the chosen parameters, before (3) the system may go first through a chaotic region~\cite{Ruiz2020}. Indications of this can be observed in Figs.~\ref{Fig3}(A) and (B) between  $p \sim 1.7$ and $p \sim 2.4$.  Comparing Figs. \ref{Fig3}(A) and (B) we find that the presence of the phonons does not change the orbits and different regimes qualitatively: the most notable effect is the stabilization of the LC and PD regimes (note in Figs.~\ref{Fig3}(A) and (B) the more extended region were PD is observed when the mechanics is present). It is a notable fact that the  frequency of the LC's and the PD orbits are set by the phonon frequency.  Indeed, the Larmor precession is subject to intensity noise, because $\bm{B}$ depends on the polariton population. The mechanical clock consequently stabilizes the dynamics by making the frequencies constant and independent of $p$, thus facilitating the observation of the whole phenomenon.

 \paragraph*{\textbf{Discussion and outlook}} 
 
The dynamics of the reservoir, and specifically of the population imbalance $m$, is crucial for the observation of the discussed complex dynamics. Only when the reservoir population is left to respond in time to that of the trapped-polariton dynamics, for parameter values that are physically reasonable, the LCs and PD emerge. It turns out that the reservoirs' polarization $m$ is only different from zero if there is a spontaneous symmetry breaking of the condensate polarization leading to $S_z\neq0$ ($m \simeq -\frac{S_z}{2}$). 
It is this spontaneous symmetry breaking that leads to the pitchfork bifurcation in Fig.~\ref{Fig3}(C), and subsequently to the instability of the two SFP leading to the self-induced Larmor precession. Note also that the phase of the LCs and PD orbits described above is fully arbitrary. This is an important requirement for time crystalline behavior, naturally satisfied in our system which is only driven non-resonantly, with a continuous wave optical excitation, and with excitons loosing any reference to this initial conditions through relaxation to the reservoir. The second strong requirement for a time crystal is the robustness against large variations of the relevant physical parameters. That this is indeed the case is shown in Fig.~\ref{Fig3}(D), where a phase diagram for the stability of the LC+PD dynamics (only PD) is shown in green (blue) as a function of the complex coupling $J + i\gamma_d$. Note the large parameter regions where stable self-induced dynamical behavior exists, demonstrating the robustness of the limit cycles against perturbations and bringing further support for the emergence of a time crystal. The quite universal observation of similar dynamics in traps of different size and exciton-photon detuning (see section S7 of the supplementary material~\cite{SM}), further supports this conclusion. 
The qualitative agreement between the theoretical power dependencies in Fig.~\ref{Fig3}(B) and the experimental ones in Fig.~\ref{Fig2}(A-B) is also quantitative, as illustrated by characteristic spectra in Fig.~\ref{Fig3}(E), further bringing strength to the proposed model. Finally, the model is also predictive in the sense that, as follows from Fig.~\ref{Fig3}, a forced mechanical driving (instead of the self-induced one described so far) should also lead to PD behavior. This was experimentally tested using coherent mechanical phonons externally generated with pulsed lasers~\cite{Kimura2007} (see S9 of the supplementary material~\cite{SM}).


In summary, we have demonstrated the remarkable dynamics of a driven-dissipative, non-linear many-body solid-state quantum cavity polariton system coupled to confined vibrations. Fed by a fully incoherent exciton bath, this system exhibits continuous time crystalline behavior involving the spinor of the trapped condensate's ground state and its coupling to a dynamic exciton reservoir. Period-doubling occurs without any external time-dependent harmonic driving, referenced to mechanical coherent vibrations that are self-induced by the polariton fluid. These self-induced mechanics stabilize and lock the frequency of the time crystal, acting as an internal clock. Our findings pave the way for exploring time crystals in open many-body quantum systems. Moreover, the combination of non-linear photonic arrays with self-induced oscillations could serve as a testbed for dynamical gauge theories, with applications including quantum simulators for quantum electrodynamics and quantum chromodynamics models~\cite{Walter2016}.







\begin{acknowledgments}
We acknowledge partial financial support from the ANPCyT-FONCyT (Argentina) under grants PICT-2018-03255, PICT 2018-1509, PICT 2019-0371, PICT 2020-3285, and SECTyP UNCuyo 06/C053-T1.
ASK and PVS acknowledge the funding from German DFG (grant 359162958). GU thanks J. Tempere and M. Wouters for discussions during a research stay at UAntwerpen, partly funded by the
Fund for Scientific Research-Flanders and N. Goldman for his hospitality at ULB. 
\end{acknowledgments}







\onecolumngrid

\pagebreak

\setcounter{section}{0}
\setcounter{page}{1}


\renewcommand{\thesection}{S\arabic{section}}
\setcitestyle{round}

\renewcommand{\figurename}{Figure \!\!}
\renewcommand{\thefigure}{\textrm{S}\arabic{figure}}

\renewcommand{\theequation}{\textrm{S}\arabic{equation}}

\begin{center}
\textbf{\Large \underline{Supplementary Material:}\\~\\ \huge Solid-state continuous time crystal with a built-in clock}
\end{center}

\onecolumngrid  

\section{Methods and materials}
\subsection{Microcavities fabrication}
\label{Section: sample}

The hybrid photon/phonon microcavities reported in this work were grown by molecular beam epitaxy (MBE) on a double polished GaAs substrate. The layer structure consists of several layers of (Al,Ga)As forming distributed Bragg reflectors (DBRs). In between the described DBRs, a cavity spacer with embedded GaAs quantum wells (QWs) is grown in such a manner that the electrostrictive polariton-phonon interaction is optimized. To achieve this optimization, the QWs were placed in positions where the product of cavity strain and cavity electric field is maximized. This sample design results in a strong modulation of the acoustic and optical properties, leading to the concurrent confinement of phonons and photons within the cavity and maximizing their mutual coupling \cite{Chafatinos2023}.

During the MBE process, the growth is stopped after the deposition of the spacer layer and the sample is then photolitographically patterned and processed with a few nanometers wet etching.  Afterwards, the sample is reinserted into the MBE for the overgrowth of the top DBR. The engraved pattern is preserved during the overgrowth, enabling lateral trap confinement due to the difference in the acoustical and optical cavity resonance energies between the etched and non-etched regions of the sample. The overgrowth of the top DBR is performed at a lower temperature than usual to minimize the smoothing of the lateral interfaces. The detail process is well described in Ref.\,[\onlinecite{Kuznetsov2018}].

The primary sample utilized in this study (identified as sample A) features a top (bottom) distributed Bragg reflector (DBR) composed of 25 (33) $\lambda$/4 periods of Al$_{0.15}$Ga$_{0.85}$As/Al$_{0.90}$Ga$_{0.10}$As layers, embedding a 3/2$\lambda$ cavity  housing four 15 nm GaAs quantum wells (QWs) and grown on a 350 $\mu$m thick GaAs(001) substrate. This sample was specifically designed to confine longitudinal acoustic phonons with a frequency $\nu^{\lambda}_m\sim20$ GHz (additional details can be found in Ref.\,[S\onlinecite{Chafatinos2023}]).

A second sample (sample B) was employed for the experiments presented in section S9 of this supplementary material. In this case, the DBR period comprises a stack of three pairs of Al$_{x}$ Ga$_{(1-x)}$As/Al$_{y}$Ga$_{(1-y)}$As layers with an optical thickness of $\lambda$/4. As described in detail in Ref.\,[S\onlinecite{Kuznetsov2023}], this configuration induces the simultaneous confinement of photons of wavelength $\lambda$, and acoustic phonons of wavelength $\lambda$ and $3\lambda$. This results in confined acoustic longitudinal vibrations with fundamental frequencies of both $\nu^{\lambda}_m\sim21$ GHz and $\nu^{3\lambda}_m\sim7$ GHz.



\subsection{Time-resolved first order spatial correlations measurement technique}
\label{Section: g1}

\begin{figure}[t]
    \centering \includegraphics[width=0.6\linewidth]{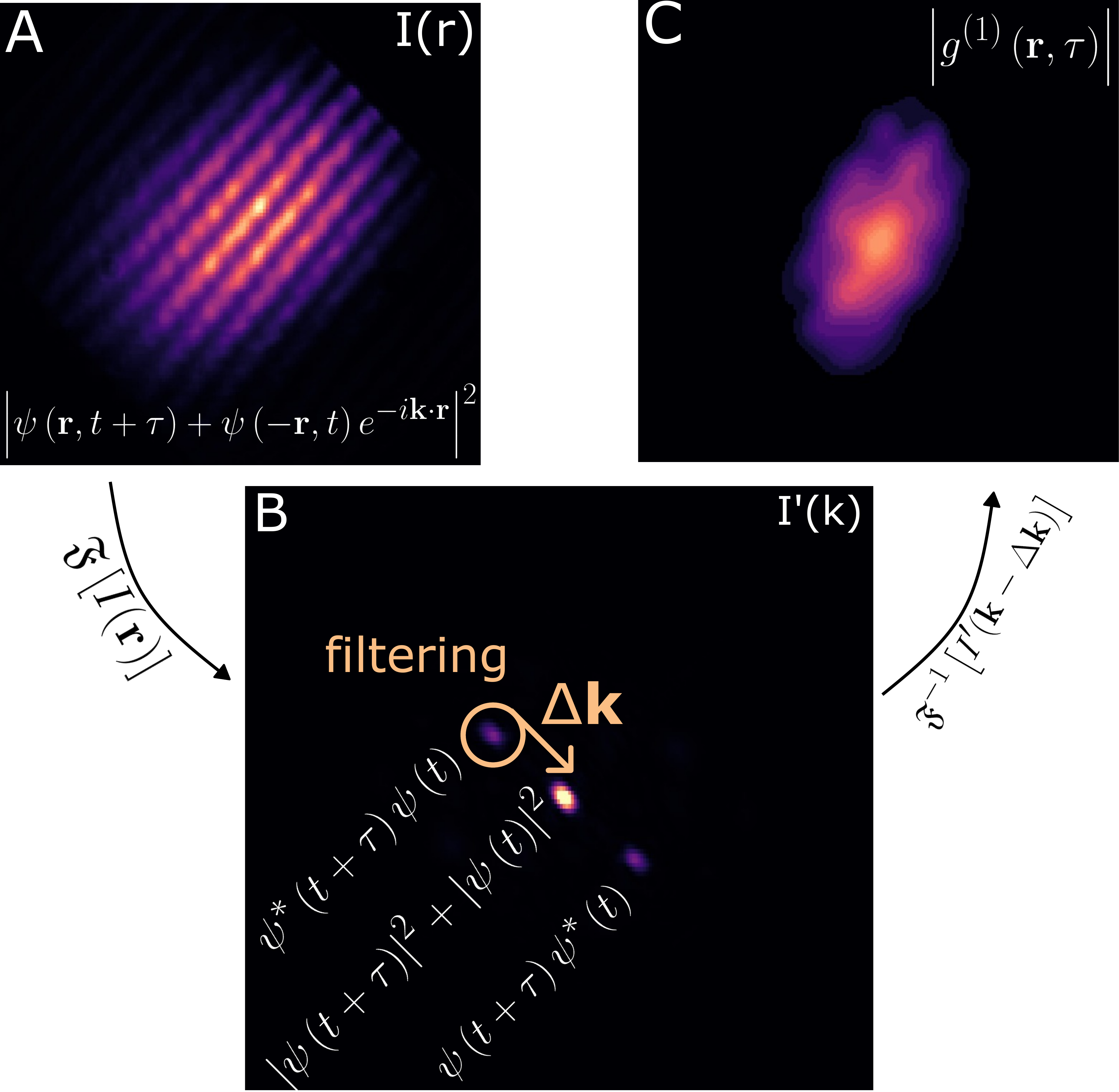}
    \caption{\textbf{A} Interference pattern generated by the emitted signal from a square trap. \textbf{B} Application of a two-dimensional Fourier transform to the interference pattern reveals three intense peaks. The two peaks located at $\pm\Delta\mathbf{k}$ contain information on the correlation between delayed and non-delayed beams. To extract this information, a displacement by $-\Delta\mathbf{k}$ is applied, followed by a filtering process targeting one of these peaks. \textbf{C} $g^{(1)}$ in real-space is obtained through the inverse Fourier transform of \textbf{B}.}
    \label{f: figg1}
\end{figure}

The time-resolved spatial first-order coherence function, denoted as $g_{i,j}^{(1)}(r, -r, \tau)$, represents the normalized correlation between one field at point $r$ with itself ($i=j$) or another field ($i\neq j$) at point $-r$ and observed at a delayed time $\tau$. Mathematically, it is expressed as:

\begin{equation}
    g_{i,j}^{(1)}(r,-r, \tau) = \frac{\left<\psi_i^*(r,t+\tau) \psi_j(-r,t)\right>}{\sqrt{\left<\left|\psi_i(r,t+\tau)\right|^2\right>\left<\left|\psi_j(-r,t)\right|^2\right>}},
\end{equation}
where $<\cdot>$ denotes the ensemble average taken over all spatial and temporal points. The field, denoted as $\psi_i(r,t)=A_i(r,t)e^{-i\phi_i(r,t)}$, is described by an amplitude $A$ and a phase $\phi$ of the signal emission at a specific spatial point $r$ and time $t$. We are interested in the correlation of the signal with itself, thus $i=j$ and $g^{(1)}(r,-r,\tau) = g_{ii}^{(1)}(r,-r,\tau)$. Since the correlation function was measured for the ground state of the trap, which has spatial-inversion symmetry, we can reduce the function to $g^{(1)} (r,\tau) = g^{(1)} (r,-r,\tau)$. The measurement of $g^{(1)}$ is performed using a modified Michelson interferometer with a mirror-retroreflector configuration as illustrated in Fig.~2C of the main text, where the displacement of the mirror allows for the adjustment of the temporal delay $\tau$. To derive the value of $g^{(1)}$ we performed a Fourier filtering procedure on the measured interference patterns, as  schematically represented in Fig.~\ref{f: figg1}. The emitted signal from the trap is split into two optical beam paths --one with a delayed trajectory and the other undergoing spatial retro-reflection-- and are made to interfere on the CCD detector. Figure~\ref{f: figg1}A displays the real-space interference pattern obtained for an experiment at a specific delay time $\tau$. The detected intensity is given by $\left | \psi(\mathbf{r},t+\tau)+\psi(-\mathbf{r},t) e^{-i\mathbf{\Delta k\cdot r}} \right |^2$, where the term $\Delta\mathbf{ k\cdot r}$ is determined by the angle between the beams at the entrance slit of the spectrometer.

Once acquired, the interference patterns are processed as follows: First a two-dimensional Fourier transform is performed, enabling the discrimination of different spatial frequencies. The resulting Fourier transform in Fig.~\ref{f: figg1}B reveals three peaks. The central peak corresponds to the term $\left | \psi(t+\tau)\right |^2 + \left |\psi(t) \right |^2$, and contains information about each independent beam. The two adjacent peaks, translated by $\pm \Delta\mathbf{k}$ relative to the origin, arise from the products $\psi^*(t+\tau)\psi(t)$ and $\psi^*(t)\psi(t+\tau)$, providing information about the correlation between the beams.

To extract the correlation information, we first apply a translation equal to $-\Delta \mathbf{k}$, then a filtering process selectively retains frequencies within a small region at the center, eliminating the other peaks in the reciprocal space. Afterwards, the inverse Fourier transform of the filtered momentum space image yields $|g^{(1)}(r,\tau)|$, as presented in Fig.~\ref{f: figg1}C. Finally, a spatial integration is performed to obtain a data point for the delay time $\tau$, as shown in Figure 2D of the main text.


\section{Gross-Pitaevskii equation for the polariton spinor states}
A polariton condensate is a collective many-body state formed by excitons and photons. In the simplest (mean field) approximation, such a state of matter can be described by a single spinor wavefunction, $\Psi=(\psi_+,\psi_-)^\mathrm{T}$. The pseudo-spin degree of freedom originates from the coupling between the excitons' spin and the microcavity photons' polarization, thus it can be measured through the polarization of the emitted light.  One can model the pseudospin's dynamics through the following set of coupled generalized Gross-Pitaevskii equations (gGPEs) \cite{CarusottoRMP2013},
\begin{eqnarray}
\nonumber
i\hbar\dt{\psi_\sigma} &=&(\varepsilon_\sigma+U_1|\psi_\sigma|^2+U_2|\psi_{\bar{\sigma}}|^2+U^R n_\sigma)\psi_\sigma\\
&&-(J+i\hbar\gamma_d)\psi_{\bar{\sigma}}+\frac{i\hbar}{2}(Rn_\sigma-\gamma)\psi_\sigma\,,
\label{eq: pm}
\end{eqnarray}
and the corresponding ones for the exciton's reservoirs densities
\begin{equation}
\dt{n_\sigma}=P_\sigma-\left(\gamma_R+R|\psi_\sigma|^2\right)n_\sigma+\gamma_s (n_{\bar{\sigma}}-n_\sigma)\,.    
\end{equation}
Here $\sigma=\pm$ describes the two circularly polarized states of the photon along the growth axis of the microcavity.
In the above equations, $\varepsilon_\sigma$ is the bare energy of the $\sigma$-mode, and $U_1$ ($U_2$) accounts for the same-spin (cross-spin) repulsive ($U_1,\ U_2 > 0$) Coulomb interaction between polaritons. The complex coupling $-(J+i\hbar\gamma_d)$ contains two contributions: (i) a Josephson-like coupling ($-J$), that arises from the energy splitting of the $X$ and $Y$ linearly polarized states (see discussion below); (ii) a dissipative coupling  ($-i\hbar\gamma_d$)  originated on the difference between the effective decay rates of the $X$-$Y$ linear modes (see Fig. \ref{f: figExpSM1}). 
Finally, the last term represents the stimulated gain from the exciton's reservoir (term proportional to $R$) and the bare polariton decay rate ($\gamma$).
The dynamics of the reservoirs is controlled by the pump power ($P_\sigma$), the exciton's non-radiative decay rate ($\gamma_R$) and the stimulated decay to the condensate. A spin-flip rate $\gamma_s$ between the reservoirs is also included.
Different limits of the model presented in Eq.~\eqref{eq: pm} have been extensively used in the literature to explain a diversity of phenomena related to synchronization, weak lasing and spinor dynamics \cite{Wouters2008}(see also Refs.\,[S\onlinecite{Eastham2008}, S\onlinecite{Ohadi2018}]). 

For a better analysis of the different regimes, it is useful to make the following renormalization: $\tilde{\psi}_\sigma=\psi_\sigma/\sqrt{\rho_0}$, and  $\tilde{n}_\sigma=n_\sigma/n_0$ where we introduced the densities $n_0=\gamma/R$ and $\rho_0=\gamma_R/R$. Assuming for simplicity $U_2=0$ (only the condition $U_1-U_2>0$ is relevant, while typically $U_1\gg U_2$) we have
\begin{eqnarray}
\label{eq:renormalized_plus}
\nonumber
i\hbar\dt{\tilde{\psi}_\sigma} &=&(\varepsilon_\sigma+U_0|\tilde{\psi}_\sigma|^2+U^R_0 \tilde{n}_\sigma)\tilde{\psi}_\sigma-(J+i\hbar\gamma_d)\tilde{\psi}_{\bar{\sigma}}\\
&&+\frac{i\hbar\gamma}{2}(\tilde{n}_\sigma-1)\tilde{\psi}_\sigma\,,\\
\nonumber
\dt{\tilde{n}_\sigma}&=&\gamma_R\left(p_\sigma-\left(1+|\tilde{\psi}_\sigma|^2\right)\tilde{n}_\sigma+\frac{\beta}{1-\beta} (\tilde{n}_{\bar{\sigma}}-\tilde{n}_\sigma)\right)\,,\\   
\end{eqnarray}
where $p_\sigma=P_\sigma/P_\mathrm{th}$, $P_\mathrm{th}=\gamma\gamma_R/R$ is the threshold power, $U_0=U_1\rho_0$, $U_0^R=U^Rn_0$, $\bar{\sigma}=-\sigma$ and $\gamma_s/\gamma_R=\beta/(1-\beta)$ with $0\le\beta<1$.
\subsection{The X-Y base}
While the $\psi_\pm$ basis is convenient to describe the polaritons since circularly polarized photon states selectively couple to specific spin exciton states, it is also interesting to write down the gGPEs in the linearly polarized $X$-$Y$ base. This is particularly relevant here due to the symmetry of our traps (square or tetragonally distorted square traps). We thus define $\psi_X = (\tilde{\psi}_+ + \tilde{\psi}_-)/\sqrt{2}$ and $ \psi_Y=i\bar{\psi}_Y= (\tilde{\psi}_+ - \tilde{\psi}_-)/\sqrt{2}$, so that the gGPEs take the form
\begin{eqnarray}
\nonumber
i\hbar\dt{\psi_a} &=&\left[\varepsilon_a+\frac{1}{2}U_0\left(|\psi_a|^2+|\psi_b|^2\right)+U^R_0 n\right]\psi_a\\
\nonumber
&&+\left[J'+\left(U^R_0+\frac{i\hbar\gamma}{2}\right)m+U_0\, \mathrm{Re}(\psi_a^*\psi_b)\right]\psi_b\\
&&+\frac{i\hbar\gamma}{2}\left(n-\frac{\gamma_a}{\gamma}\right)\psi_a\,,
    \label{eq:renormalized_X}
\end{eqnarray}
with $a=X,Y$ ($b=Y,X)$, $\varepsilon_{X,Y}=(\varepsilon_++\varepsilon_-)/2\mp J$, $J'=\Delta\varepsilon/2=(\varepsilon_+-\varepsilon_-)/2$, $\gamma_{X,Y}=\gamma\pm2\gamma_d$, $n=(\tilde{n}_++\tilde{n}_-)/2$ and $m=(\tilde{n}_+-\tilde{n}_-)/2$. Note that $\gamma_X<\gamma_Y$ for $\gamma_d<0$, and therefore condensation is expected to take place on the $X$ polarized mode (which is the lowest energy state if  $J>0$).

With this change of variables some important points become clearer.
Firstly, the dissipative coupling in the $(+,-)$ spinor base emerges from a difference in the decay rate of the $X$ and $Y$ modes whereas in the $X$-$Y$ basis it arises from the spin imbalance of the circularly polarized reservoirs, being proportional to $i\hbar\gamma m/2$.   
Secondly, there are two terms that couple both modes which arise from the interactions: i) one proportional to $U_0\mathrm{Re} \left( \psi_X^* \psi_Y\right)$, which depends on the relative phase of the modes; ii) and the other proportional to $U_0^R m$, that depends on the (circularly polarized) pseudospin imbalance of the reservoirs. 
Both of these terms will be relevant for the self-oscillating dynamics in which we are interested. Lastly, we note that the Josephson-like coupling in the $(+,-)$ basis appears as a bare energy splitting in the $X$-$Y$ basis. Conversely, the bare splitting of the $(+,-)$ states, ($\varepsilon_+ - \varepsilon_-$), corresponds to the Josephson-like coupling between $X$-$Y$.
\subsection{Amplitude-phase description}
There is yet another alternative description in terms of the occupation and the phase of the modes that allows for a more direct link between coupled polariton and superconducting junction phenomena. Using the Madelung transformation, $\tilde{\psi}_\sigma = \sqrt{N_\sigma} e^{i \theta_\sigma}$ in Eq. \eqref{eq:renormalized_plus} we obtain
\begin{equation}
 \dt{N_\sigma}=\gamma  \left(\tilde{n}_\sigma-1\right) N_\sigma+2 \sqrt{N_+ N_-} \left(\sigma \frac{J}{\hbar} \sin(\theta)-\gamma_d \cos
   (\theta)\right)\,,
   \label{eq: N+}
\end{equation}
and 
\begin{eqnarray}
\nonumber
\hbar\dt{\theta}&=& (\varepsilon_--\varepsilon_+)+(N_--N_+)U_0+(\tilde{n}_--\tilde{n}_+)U_0^R\\
\nonumber
&&+\frac{\left(N_--N_+\right) }{\sqrt{N_+ N_-}}J \cos(\theta)+\frac{ \left(N_++N_-\right) }{\sqrt{N_+ N_-}}\hbar\gamma_d\sin(\theta)\,.\\
       \label{eq:theta}
\end{eqnarray}
Here, $\theta = \theta_+ - \theta_-$ is the phase difference between the two circularly polarized polariton modes.
The first term in Eq.~\eqref{eq: N+} governs the net gain from the reservoirs and describes a Lotka-Volterra type of competition when the spin flip rate in the reservoirs is non-zero ($\beta\ne0$)~(see Ref.\,[S\onlinecite{Lotka-Volterra}]). 
The second term, proportional to $J$, depends on the relative phase $\theta$ and thus can be directly related to a Josephson current, analogous to the one present in superconducting junctions. We recall here that this (Josephson-like) coupling is due to the energy splitting between the $X$-$Y$ modes. The third term has a similar structure as the second, and represents an additional ``dissipative current'' between the modes, arising from the dissipative coupling $\gamma_d$. Note that this new contribution is out of phase with the Josephson current.

We now focus on the relative phase presented in Eq.~\eqref{eq:theta}. Equivalently to the Josephson equation derived for superconducting junctions, $\theta$ grows with the energy splitting between the $+$ and $-$ modes (which plays the role of a voltage). However, differently from standard superconductors, it is renormalized by the interactions. The last two terms are the ones that couple the population and phase through the previously described currents. Note that the dissipative current term is proportional to the global population, ($N_+ + N_-$). This is in contrast to conservative currents, where only the relative population is relevant.

The same transformation can be applied to the $X$-$Y$ base by introducing the notation $\psi_a = \sqrt{N_a} e^{i \theta_a}$ in Eq.\,\eqref{eq:renormalized_X}, 
\begin{eqnarray}
\nonumber
\hbar \dt{N_a}&=&\hbar\gamma\left(n-\frac{\gamma_a}{\gamma}\right)N_a -\xi_aU_0 N_XN_Y \sin(2\phi) \\
\nonumber   
&&+\sqrt{N_X N_Y}\,\hbar\gamma m \cos (\phi )\,,\\
&&-2\xi_a\sqrt{N_X N_Y}\left(J'+m U_0^R\right)\sin (\phi)\,, 
\label{eq: NX}
\end{eqnarray}
with $\xi_X=-\xi_Y=1$, $\phi=\theta_X-\theta_Y$, and
\begin{eqnarray}
\nonumber
\hbar\dt{\phi}&=&2J+U_0 \cos ^2(\phi )(N_X-N_Y)\\
\nonumber
    &&+\frac{\left(N_X-N_Y\right)}{\sqrt{N_X N_Y}}\left(J'+m U_0^R\right)\cos (\phi ) \\
&&-\frac{\left(N_X+N_Y\right)}{\sqrt{N_X N_Y}}\frac{\hbar\gamma}{2}  m \sin (\phi )\,.
    \label{eq: phi}
\end{eqnarray}
Similarly to the $(+,-)$ case, the first term of Eq.~\eqref{eq: NX} determines the net gain.  
Notice however, that since this term depends only on $n$ for both polarizations, no Lotka-Volterra competition emerges.
The terms proportional to $\xi_a$  correspond to conservative currents, and arise from the interaction between polaritons, the interaction with the reservoir (notice the different phase dependence of each of them),  and a Josephson contribution that steams from the coupling $J'$.  The third term corresponds to a  ``dissipative current'' which depends on the pseudospin imbalance throughout $m$. 
Equation~\eqref{eq: phi} for the phase difference reflects all these contributions, where the contribution associated to the dissipative current is again proportional to the global population, $N_X+N_Y=N_+ + N_-$.

In brief, we find that the difference of dissipation plays a central role in the evolution of the pseudospin. This last phenomena can be induced through a direct linewidth difference as is the case for the $X$ and $Y$ modes through $\gamma_d$, or through a population imbalance between the two reservoirs. Note also how the role of (diagonal) energy splitting and (non-diagonal) mode-coupling are interchanged when the representation is changed between the circularly $(+,-)$ and linearly $X$-$Y$ polarized bases.

\section{Spinor description: the Poincare sphere}
The Gross-Pitaevskii and amplitude-phase set of equations presented above are useful to make the connection with related dynamics observed in superconductor junctions, coupled atomic condensates, and arrays of non-linear lasers. Synchronization, Josephson oscillations and self-trapping, are some of the solutions that emerge from these equations. The different terms, involving driven-dissipative conditions, reservoir competition, complex inter-state couplings, and non-linearities, lead to a complex dynamics that is dependent upon the limits considered. An alternative approach that might be suited to grasp the physics expected from the different contributions, and provides a good starting point to consider the effect of a time-dependent mechanical modulation, derives from a spinor description in the Poincare sphere. We therefore change to a spin formalism with a pseudospin vector defined as $S_j= \tilde{\Psi}^\dagger \sigma_j \tilde{\Psi}$, where $\sigma_j$ are the different Pauli matrices. Based on this definition, each component of the pseudospin takes the form $S_x = 2 \mathrm{Re} \left( \tilde{\psi}_+\tilde{\psi}_-^* \right)=2 \cos(\theta)\sqrt{N_+N_-}$, $S_y = 2 \mathrm{Im} \left( \tilde{\psi}_+ \tilde{\psi}_-^* \right)=2 \sin(\theta)\sqrt{N_+N_-}$ and $S_z = \left| \tilde{\psi}_+ \right|^2 - \left| \tilde{\psi}_- \right|^2=N_+-N_-$. This transformation has a direct physical interpretation through the polarization of the emitted light. The polarization of the photon component of the polaritons maps directly with their pseudospin. 
In this sense the magnitude and sign of $S_x$, $S_y$ and $S_z$ is just a measurement of how the light is horizontally, diagonally or circularly right polarized (vertical, anti-diagonal or circularly left, if the components are negative). 
Using this notation we obtain the following set of equations for the dynamics of the pseudospin components,
\begin{equation}
    \dt{\bm{S}}=\frac{1}{\hbar}\bm{B}\times\bm{S}+S\bm{E}+ \Gamma\bm{S}\,,
        \label{eq: S-SM}
\end{equation}
where
\begin{eqnarray}
\nonumber
\bm{B}&=&2J\,\xv-(\Delta\varepsilon+U_0S_z+2U_0^Rm)\,\zv\,,\\
\nonumber
\bm{E}&=&-2\gamma_d\,\xv+\gamma m\,\zv\,,\\
\Gamma&=&\gamma(n-1)\,,
\label{eq: BE-SM}
\end{eqnarray}
and, as before, $n=(\tilde{n}_++\tilde{n}_-)/2$ and $m=(\tilde{n}_+-\tilde{n}_-)/2$. In terms of the spin variables these latter quantities satisfy
\begin{eqnarray}
    \label{eq: nm}
    \dt{n}&=&\frac{1}{2} \gamma _R \left(p_++p_--m S_z-n(S+2)\right)\,,\\
    \nonumber
    \dt{m}&=&\frac{1}{2} \gamma _R \left(p_+-p_--m \left(2\frac{ (1+\beta)}{1-\beta}+S\right)-nS_z\right)\,.
\end{eqnarray}
Equation~\eqref{eq: S-SM} resembles the Bloch equations used to describe spin dynamics in magnetic resonance phenomena, albeit with some important differences. Starting with the similarities, note that the first term with the vector product is the one leading to Larmor precession and all the coherent control strategies using pulsed resonant fields. The last term is similar to the phenomenological way in which spin decay times are usually introduced, where $\Gamma$ describes the net gain of the system. The second term could be associated with a synthetic crystal-field $\bm{E}$ that favors specific orientations for the spinors.

The synthetic magnetic field $\bm{B}$, crystal field $\bm{E}$ and relaxation rate $\Gamma$ involve complex non-linear dependencies that arise from the interactions and the reservoirs suppression by the condensate (feedback terms proportional to $S$ and $S_z$ in Eq.\,\eqref{eq: nm}). $\bm{B}$ has two fixed components, one in the $x$ direction, originated in the Josephson coupling $J$ (i.e., in the energy splitting of the $X$-$Y$ modes), and another along $z$ represented by the pseudospin splitting $\Delta\varepsilon$ of the $(+,-)$ modes. This latter contribution is  equivalent to a Zeeman term, and is only different from zero in the presence of an external magnetic field. But, note that $\bm{B}$ also has a non-linear component in the $z$ direction that changes with the population imbalance of the $(+,-)$ modes ($S_z$) and that of the reservoirs ($m$). This implies that, in general, the Larmor precession leads to a self-induced time dependence in the magnetic field. The second term in Eq.~\eqref{eq: S-SM} contains the field $\bm{E}$ whose $x$ component reflects the coupling caused by the difference of dissipation between the $X$ and $Y$ modes, and the $z$ component derives from the difference of dissipation between the $(+,-)$ modes caused by the reservoir imbalance $m$. Again, this second term involves a non-linearity, and thus can be time-dependent if a self-sustained dynamics sets in.

At this point, it is insightful to make the usual approximation that the reservoirs are nearly stationary, that is $dn/dt=dm/dt=0$, so that 
\begin{eqnarray}
\nonumber
    n&=&\frac{2(2+S)p}{4+4S+S^2-S_z^2}\,,\\
    m&=&-\frac{2S_z p}{4+4S+S^2-S_z^2}\,,
    \label{eq: nm_instant}
\end{eqnarray}
where, for the sake of simplicity, we assumed for hereon $p_+=p_-=p$ (i.e., excitation with unpolarized or linearly polarized light) and $\beta=0$ (i.e., no cross feeding from the reservoirs with different spins). Furthermore, above the threshold power, we can take $p=1+\delta p$ with $0\le\delta p$ (up to small corrections $\sim\gamma_d/\gamma\ll1$ due to renormalization of $P_\mathrm{th}$, see below). Since all the pseudospin components are proportional to $\delta p$, to linear order in $\delta p$  we have
\begin{eqnarray}
\nonumber
    n&=&1+\delta p-\frac{S}{2}\,,\\
    m&=&-\frac{S_z}{2}\,
\end{eqnarray}
Note that the reservoirs' polarization $m$ is only different from zero if there is a spontaneous symmetry breaking of the condensate polarization leading to $S_z\neq0$. In this approximation we then have
\begin{eqnarray}
\label{low_power_fields}
\nonumber
\bm{B}&\sim&2J\,\xv-(\Delta\varepsilon+(U_0-U_0^R)S_z)\,\zv\,,\\
\nonumber
\bm{E}&=&-2\gamma_d\,\xv-\gamma \frac{S_z}{2}\,\zv\,,\\
\Gamma&=&\gamma\left(\delta p-\frac{S}{2}\right)\,.
\end{eqnarray}
Note that there is a competition between the polariton-polariton and polariton-reservoir interactions, the latter being effectively attractive, and which results in an effective interaction $U=U_0-U_0^R$. We would like to point out, that these simple equations are only valid near $P_\mathrm{th}$, but the different role (sign) of $U_0$ and $U_0^R$ remains valid at higher powers.

In the following section we will discuss the different contributions to the spin dynamics described by Eq. \eqref{eq: S-SM}. For convenience, from hereon, all energies will be  expressed in units of $\hbar\omega_0$ and the rates in units of $\omega_0$, where $\omega_0$ is some characteristic angular frequency (it will be associated later on to the frequency of the confined phonon).
\section{Self-sustained driven-dissipative spinor dynamic}
\subsection{Larmor precession}
\begin{figure}[t]
    \centering \includegraphics[width=0.6\linewidth]{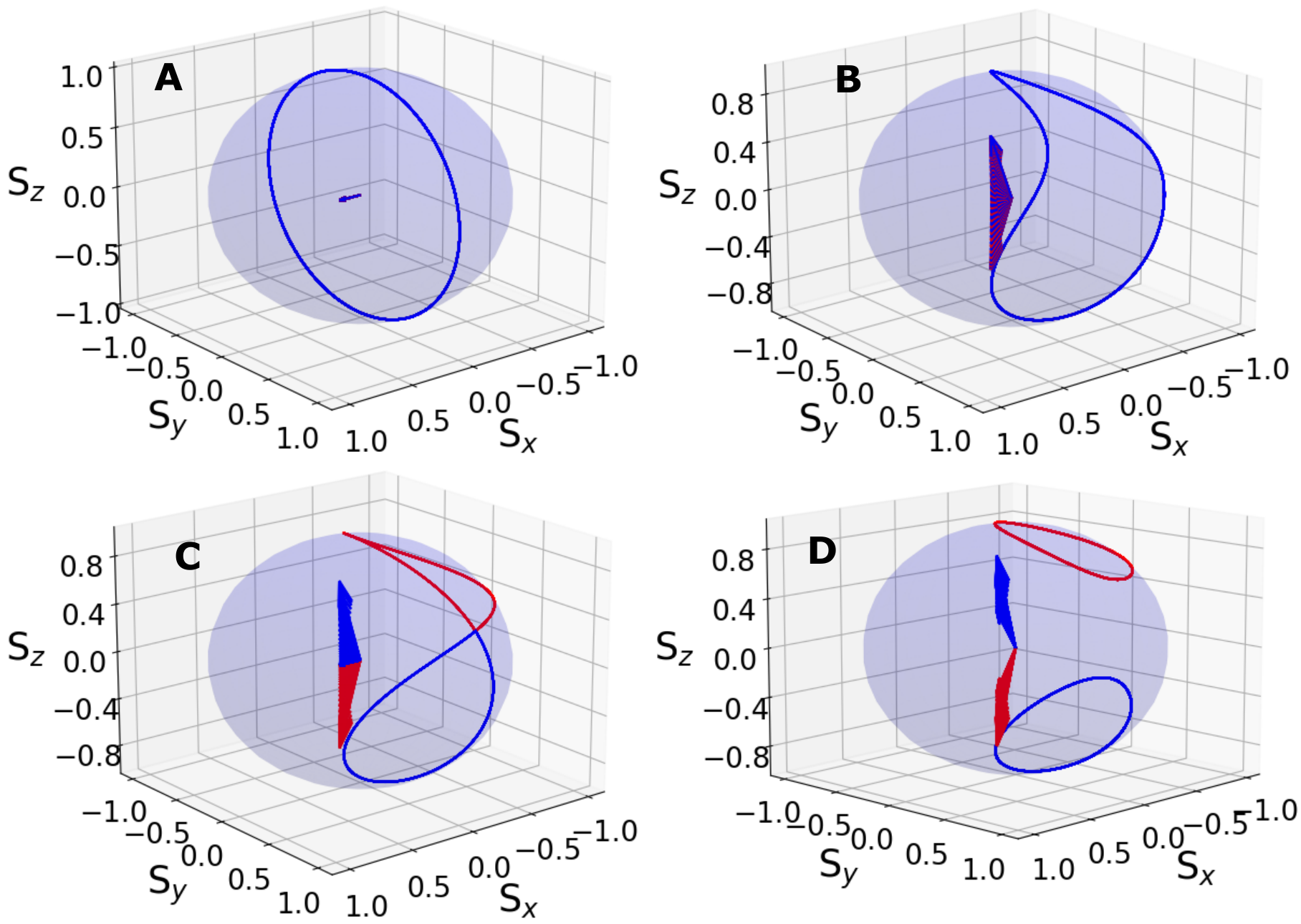}
    \caption{Poincare spheres depicting different NMR-like trajectories obtained from Eq. \eqref{eq: S-SM} using the approximated fields of Eq. \eqref{low_power_fields} without the driven-dissipative terms ($\gamma=\gamma_d=0$). The effective magnetic field takes the form $\bm{B} = 2J \hat{x} - U S_z$, with $J = 0.125$. Panels \textbf{A, B, C} and \textbf{D} corresponds to $U = 0, 0.35, 0.5, 0.55$, respectively. The different colors of each orbit represent the different initial conditions, $\bm{S}_0 = \pm\hat{z}$ (red/blue), showing different attractors. The arrows depict the effective magnetic field $\bm{B}$ following the same color scheme. The oscillating pattern of the arrows in panels \textbf{B, C)} and \textbf{D} represents the self-induced periodic dynamics of the field.}
    \label{f: figSM1}
\end{figure}

We begin the analysis for the case with no dissipation. That is, when the last two terms of Eq.~\eqref{eq: S-SM} are equal to zero ($\gamma=\gamma_d=0$). If there are no interactions ($U_0=U_0^R = 0$), the pseudospin simply precesses around the synthetic magnetic field $\bm{B}=2J\,\xv-\Delta\varepsilon\,\zv$, as expected from the regular Bloch's equations.  
Figure \ref{f: figSM1}~A shows this Larmor precession in the Poincare sphere where the arrow represents the magnetic field. Here we assumed $\Delta\varepsilon=0$, relevant for the experimental situation---a finite $\Delta\varepsilon$  just tilts $\bm{B}$ and leads to similar trajectories. The circular orbits on the Poincare sphere are determined by the initial conditions as $\bm{S}(t)\cdot\bm{B}=\bm{S}_0\cdot\bm{B}$ is a constant of motion.

The situation changes when non-linearities ($U \neq 0$) are added. Here, for simplicity, we used the form of $\bm{B}$ described in Eq.~\eqref{low_power_fields}. Figure~\ref{f: figSM1}~B shows the corresponding (overlapping) orbits  for the same  initial conditions as in panel A ($\bm{S}_0 = \pm\hat{z}$). In this case, the trajectory is distorted due to the oscillating magnetic field induced by the Larmor precession coupled to the non-linearity. This is evidenced by the oscillating pattern of the arrows representing the magnetic field that are modulated by the term $US_z$ in the $z$ component. The self-induced time-dependent change in the magnetic field orientation determines the instant axis around which the Larmor precession evolves.

Something interesting happens when $U$ is large enough, as displayed in Fig.~\ref{f: figSM1}~C. Close to $U\sim0.5$ (for $J=0.125$) there is a bifurcation in the system where the mean value of $\left< S_z \right>$ spontaneously breaks from zero. This is evidenced by the appearance of two orbits, with opposite values for $\left< S_z \right>$, that depend on the initial conditions, here given by $\bm{S}_0 = \pm\zv$ (red/blue colors in the figure). This means that for large enough $U$ there is a spontaneous symmetry breaking (SSB) for the solutions of the system where they acquire a non-zero circularly polarized component. 
We find that these new solutions become more polarized as the value of $U$ becomes larger (as shown in Fig.~\ref{f: figSM1}~D), and where the orbits are more separated. In the following subsections we show that a similar SSB of the solutions survives when the non-linear dissipative terms are turned on.

\subsection{The role of dissipation} \label{sec4 - subsecB}
When dissipation is included the dynamics of the system changes drastically and presents a complex behavior as a function of the parameters, and strongly dependent on the pump power. In terms of the spin variables, this essentially controls the value of $S$ and hence the relevance of the interaction and reservoir's suppression terms.

\begin{figure}[t]
    \centering \includegraphics[width=0.4\linewidth]{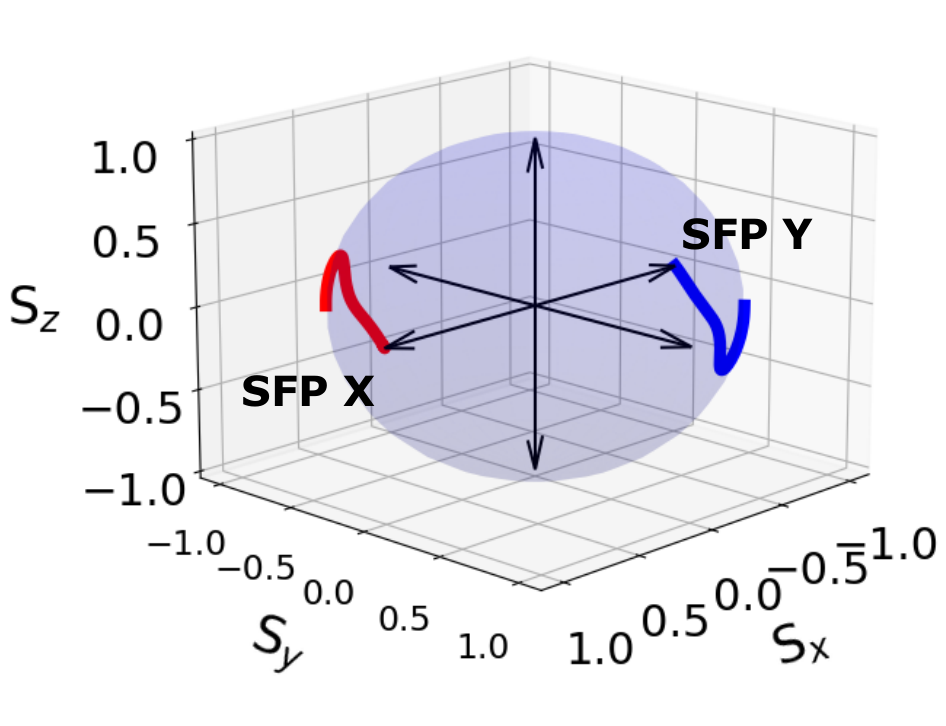}
    \caption{Trajectories on the normalized Poincare's sphere for the solutions of Eq. \eqref{eq: S-SM} without interactions ($U_0=U_0^R=0$), for a magnetic field $\bm{B} =  2J \xv$, with $2J=0.45$, a crystal field $\bm{E} = -2\gamma_d \xv+\gamma m \zv$, where $\gamma = 0.85$, $\gamma_R = 0.4$, $2\gamma_d = -0.05$, and $p = 2.5$. We plot two different orbits with initial conditions $S_0 =\pm 1.4 \xv \pm 1.4 \yv$ in red/blue colors. SFP Y and SFP X correspond to the two possible stable solutions.}
    \label{f: figSM2}
\end{figure}

The simplest effect is the emergence of fixed points (FP), $d\bm{S}/dt=0$, that control the asymptotic dynamics. Due to the symmetry of Eqs. \eqref{eq: S-SM} and \eqref{eq: BE-SM} when $p_+=p_-$ and $\Delta\varepsilon=0$, the fixed points come in pairs. This is evident, since when $\bm{S}$ is a FP, so is $\bm{S}'=\mathcal{R}_x(\pi)\bm{S}$, where $\mathcal{R}_x(\pi)$ is a $\pi$ rotation around the $x$ axis.
Analytical expressions for such points are hard to obtain and the equations involved are rather cumbersome. Yet, there is a special case that is particularly simple:  $\bm{S}=\pm S\xv$. In this a case, $\bm{B},\bm{E}\parallel \bm{S}$ and so $m=0$ and $\Gamma=\pm2\gamma_d$, give the condition for
\be
S=2(\tilde{p}_\pm-1)\,,
\ee
with $\tilde{p}_\pm=P/P^{X,Y}_\mathrm{th}$, a new renormalized pump power, and $P^a_\mathrm{th}=\gamma_R\gamma_a/R$ the power threshold for the $X$ and $Y$ polarization modes. We refers to these fixed points (when stable) as SFP X and SFP Y, respectively.
It is then clear that if $\gamma_d<0$ ($\gamma_X<\gamma_Y$) the power threshold for the $X$ mode is smaller and the system will condensate with such polarization. We emphasize that these particular fixed points are present regardless of the presence or absence of interactions. In the later case, however, they are not always stable and other fixed points dominate the dynamics as we shall see in the next subsection.
Figure \ref{f: figSM2} shows the spin trajectories for two different initial conditions (with $U_0=U_0^R=0$) above the condensation power for both SFP X and SFP Y: depending on how close the initial conditions are to each SFP, the pseudospin has a stationary state with the same (SFP X, blue curve) or opposite (SFP Y, red curve) direction as $\bm{B}=2J\,\xv$.
The dynamics gains quite a bit of complexity by turning on the interactions, not only through the non-linear component of $\bm{B}$ but also because of the interplay of the two components of $\bm{E}$.

\subsection{Full model: spectrum and power dependence}
\begin{figure*}[t]
    \centering \includegraphics[width=.8\linewidth]{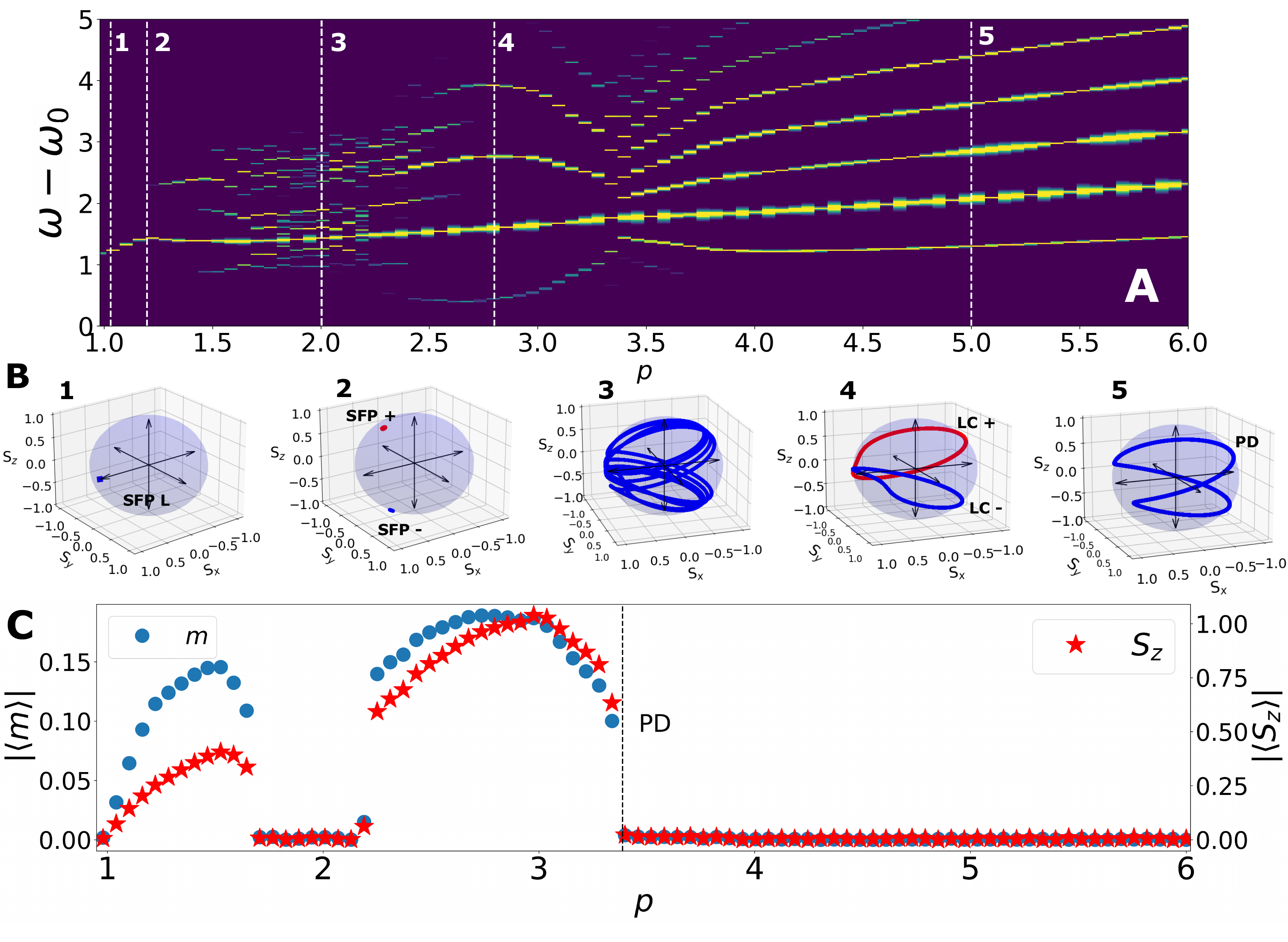}
    \caption{\textbf{A} Colormap of the energy spectrum, obtained from the Fourier transform of the solutions of Eq.\,\eqref{eq: S-SM}, plotted against the gain $p$. Simulation parameters include: $2J=0.45$, $\Delta\varepsilon=0$, $\gamma = 0.85$, $2\gamma_d = -0.05$, $U_0=0.25$, $U_0^R = 3.6$, and $\gamma_R = 0.4$, with the reservoir's dynamics governed by Eq.\,\eqref{eq: nm}. Vertical dashed white lines, each labeled with a number, indicate specific $p$ values corresponding to orbits represented on the normalized Poincare spheres on panel \textbf{B}. Red and blue colors denote distinct initial conditions, highlighting dynamical attractors. Observe the evolution of stationary solutions with $p$: beginning with a single Single Fixed Point (SFP), progressing to two SFPs, followed by a pair of limit cycle orbits (LC), and concluding with a merged orbit exhibiting period doubling (PD). \textbf{C} depicts the absolute value of the time-averaged reservoir magnetization $m$ (left axis) and the $S_z$ component (right axis) for the aforementioned solutions, as functions of the gain $p$.}
    \label{f: figSM3}
\end{figure*}

We now elaborate on the complete problem described by Eqs. \eqref{eq: S-SM}, \eqref{eq: BE-SM} and \eqref{eq: nm}, that is, incorporating dissipation, non-linear interactions, and dissipative coupling. In Fig.~\ref{f: figSM3}, various solutions for the pseudospin state are presented (see the caption for specific parameter values). Notably, non-linear interactions significantly impact the stability of the SFP X and SFP Y solutions discussed in the previous subsection. While remaining as fixed points, their stability undergoes changes, arising new attractors, both static (other SFPs) and dynamic (corresponding to limit cycles, or LCs for brevity). Fig.~\ref{f: figSM3}~A illustrates the evolution of the energy spectra of the system with increasing gain $p$. As $p$ increases, the polariton population grows, magnifying the influence of Coulomb interaction terms and the dissipative coupling terms described by the field $\bm{E}$---in this latter case a consequence of the reservoir's polarization imbalance due to the polariton induced suppression. It's worth noting that the analysis of the dependence on $p$ is particularly useful for comparison with experimental results, where the control parameter is the laser power directly linked to $p$.

In Fig.~\ref{f: figSM3}~A, we identify five distinct regimes for different values of $p$. We will describe each of these regimes.
The orbit of the pseudospin vector on the normalized Poincare sphere for five selected values of $p$ (indicated by a vertical lines and labeled 1-5 in Fig.~\ref{f: figSM3}~A) are displayed in Fig.~\ref{f: figSM3}~B).
For each case, the stationary solution for two different initial conditions with opposite $z$ component (red/blue colors) are plotted, highlighting the different possible attractors.

At low powers, the pseudospin exhibits a single SFP, denoted as SFP L (where L can be X or Y, depending on the sign and value of $\gamma_d$; in this particular case, L = X). This solution corresponds to the alignment of $\bm{S}$  with the direction a self-consistent static magnetic field on the $x$ axis (due to the $2J$ term), as illustrated by the orbit $1$ in C. This solution is essentially equivalent to what was found in the previous subsection, where interactions were not taken into account. This is to be expected, since for $p\sim 1$ one has $S\sim0$, and the non-linear terms are not effective enough against the linear terms. With an increase in gain, orbit $2$, SFP L undergoes a SSB into two new SFPs, labeled SFP+ and SFP-. These exhibit opposite $S_y$ and $S_z$ components, as they are related by a $\mathcal{R}_x(\pi)$ rotation as discussed above. It is worth noting that this SSB bears some similarities to the one discussed in the Larmor precession case: following the bifurcation, the solutions acquire a circular polarization component, breaking the symmetry along $S_z$ imposed by Eq.~\eqref{eq: S-SM}. The primary distinction from the Larmor solutions is that these are static attractors rather than dynamical orbits. Depending on the initial conditions, the condensate falls into one of these two SFPs, which are degenerate due to the $\Delta \varepsilon = 0$ condition. Therefore, to experimentally determined the obtained SFP (SFP+ or SFP-) it is sufficient to measure the polarization of the condensate. Up to this point, the obtained solutions have been static, but non-linearities in the system enable it to regain the dynamical nature that is observed in the non-dissipative case. These non-linearities play a crucial role in the formation of a driven-dissipative time crystal, which will be the focus from this point onward. 

At higher power (around $p \sim 1.2$), SFP+ and SFP- lose their stability, and the pseudospin begins to precess, as illustrated by orbit 3 (Fig.\,\ref{f: figSM3}B). Before these well-defined orbits (Larmor-like precessions) stabilize, an irregular region is formed, approximately from $p \sim 1.4$ to $2.4$. Within this region, the pseudospin traces quasi-periodic orbits in a larger portion of the sphere (orbit 3), exhibiting a dynamic reminiscent of chaos. This is evident in the spectra, where the emission lines deviate from being equally spaced (in contrast to spectra outside this region). Two microcombs form, indicating a lack of a well-defined frequency governing the dynamics. 

Around $p \sim 2.4$, the quasi-chaotic behavior disappears, and well-defined frequency orbits (evidenced by equally spaced emission lines) emerge, as depicted in orbit $4$. Two attractors persist (similar to SFP+ and SFP-), but they now take the form of limit cycles (LCs), labeled LC+ and LC-. This shift in the character of the solutions is distinctly mirrored in the spectrum presented in the upper panel. For solutions represented by orbits 1 and 2, where the attractors are fixed points, the spectrum displays a single emission line. In contrast, when the attractors are limit cycles, the spectrum exhibits two or more emission lines.

Following the formation of the LCs, with increasing $p$, the orbits of both LC+ and LC- continuously expand while their frequency undergoes changes (as evident from the varying separation between the two main emission lines in the range $p\sim2.1-3.3$). This continues until both orbits collapse into a single trajectory with a period that is clearly the double of the one of each individual orbits. The collapse into a single orbit is evident in orbit $5$ (Fig.\,\ref{f: figSM3}B). In the spectrum, the period doubling (PD) is revealed by the doubling of the emission lines at around $p\sim3.4$. Intuitively, the PD can be understood by the fact that it takes twice the time for the spinor to traverse up and down along the $S_z$ axis.

Eventually, for sufficiently large gains, this extended PD orbit collapses back to a single fixed point (SFP X or Y, depending on the sign of $\gamma_d$). While this specific outcome is not illustrated in Fig. \ref{f: figSM3}, it was already mentioned in the discussion of Fig. \ref{f: figSM2}. This observation elucidates why the condensate transitions from one linear polarization to its opposite counterpart by increasing the laser power \cite{Ohadi2018}.

Finally, we focus on the reservoir magnetization, $m$, and the circular polarization component, $S_z$. Fig. \ref{f: figSM3}~C depicts the absolute value of the time average of each of these components as a function of $p$, directly comparable with panel A. This analysis provides a clearer perspective on the spontaneous symmetry breaking observed in the SFP$\pm$ and LC$\pm$ solutions. Notably, both $\left|\left<m\right> \right|$ and $\left|\left<S_z\right> \right|$ acquire finite values in the range where these solutions exist, a crucial detail for our later discussion on the dissipation differences among emission lines.
In contrast, for the irregular or quasi-periodic solutions ($p\sim1.4-2.4$), both components approach zero. This is a consequence of the pseudospin covering the entire sphere, resulting in an average time spent in each upper and lower hemisphere, which adds up to zero. A similar behavior is observed for the PD orbit, but it follows a regular oscillatory pattern in time.

It is noteworthy that for solutions with LCs (LC+, LC-, and PD), the orbits are no longer confined to a sphere, implying that $\left|\bm{S}\right|$ is no longer constant. By examining the equations presented in subsection B, it becomes apparent that the dissipative coupling through $\gamma_d$ and $m$ permit the net population $N_X$~$+$~$N_Y$~(~$=N_- + N_+$~) to oscillate in time. This introduces a new degree of freedom to the dynamics, enabling self-sustained oscillations and the diverse phenomena described above.

\section{Optomechanical metronome \label{Sec: OM-metronome}}
The microcavity polaritons in our system couple strongly to the confined acoustic vibrations of the microcavity (cavity phonons) through deformation potential interaction, as thoroughly described in Refs.\,[S\onlinecite{Chafatinos2023, Chafatinos2020, Reynoso2022, Sesin2023}]. As a result of this coupling, the dynamics of both phonons and polaritons are deeply intertwined: phonons change the energy of the polariton modes and the coupling between them, and the back reaction force drives the phonons. A self-consistent approach for describing this complex mechanism will be addressed in subsection B. However, to gain some insight on the effect of the vibrations --that are ubiquitous to our system-- on the dynamics of the polaritons as described in the preceeding sections, we begin first our discussion by considering the case were the mechanical vibrations are assumed to be already present. This also applies to the case where the phonons are driven externally.

\subsection{Fixed phonons}
\begin{figure*}[t]
    \centering \includegraphics[width=0.5\linewidth]{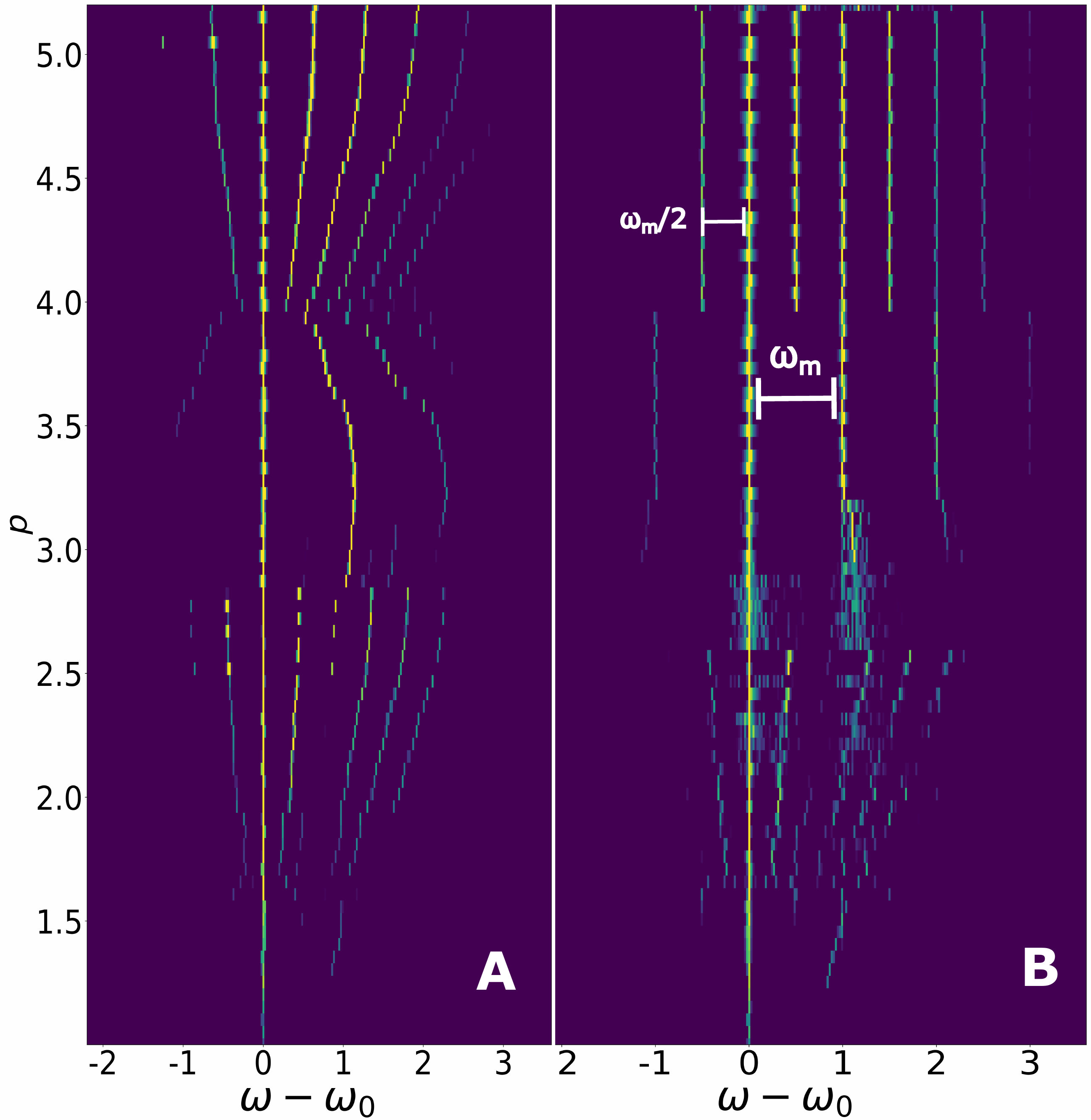}
    \caption{The energy spectrum, depicted in colormaps in panels \textbf{A} and \textbf{B}, is calculated from Eqs. \ref{eq: S-SM}, \ref{eq: BE-SM}, and \ref{eq: nm}, introducing a mechanical drive $J_p \cos{(\omega_mt)}$ to the $B_x$ component of the magnetic field. Panel \textbf{A} illustrates the scenario without phonons ($J_p=0$), while panel \textbf{B} explores the case with phonons ($J_p = 0.015$). The simulation parameters are set to $\gamma = 0.85,\ \gamma_R = 0.3,\ \gamma_d = -0.05,\ U\rho_0 = 0.5,\ U^R n_0 = 7.2, ,\ 2J_0 = 0.41$. To facilitate analysis, the blueshift is subtracted from each spectrum.}
    \label{f: forced}
\end{figure*}
One way the mechanical vibrations act on the polaritons' dynamics is by periodically modulating the hopping between two coupled states \cite{Chafatinos2023, Reynoso2022}.
In this sense, the Josephson-like coupling takes the time dependent form $J(t) = J_0 + \hbar g_0 x(t)$, where $J_0$ is the constant coupling described before, $g_0$ is the linear optomechanical coupling constant and $x(t)$ is proportional to the phonons' displacement. For a coherent phonon population we can write the displacement as \cite{Reynoso2022}, $x(t) = \sqrt{n_p} \cos{(\omega_m t)}$, $n_p$ being the number of phonons and $\omega_m$ their frequency. We then have, effectively, a oscillating $\bm{B}(t)$ field, with an $x$ component of the form: $B_x(t) = 2J_0 + 2J_p \cos{(\omega_m t)}$, where $J_p = \hbar g_0 \sqrt{n_p}$.
To analyse how this mechanical induced oscillation of $B_x$ alters the pseudospin dynamic described in previous sections we consider the case where $J_p \ll \hbar\omega_0$ as it closely resembles the experimental conditions where a relative low phonon population is present.

In Fig. \ref{f: forced}, panel \textbf{A} illustrates the dynamics without phonons ($J_p=0$), while panel \textbf{B} explores the scenario with a modest phonon contribution ($J_p = 0.015$, less than $2\%$ of the phonon energy). The color-maps in both panels present the energy spectrum as a function of gain $p$, with the energy of the most intense peak serving as a reference zero for a simplified analysis.

Despite the periodic drive in the $B_x$ field component, we observe that the qualitative features of the orbits and different regimes remain unchanged, closely resembling those presented in the normalized Poincare spheres in Fig.\ref{f: figSM3}\,B. However, a notable differecne appears in the spectrum: the frequencies of the LC's and PD orbits become locked to $\omega_m$ and $\omega_m/2$, respectively. This locking is evident in the consistent spacing between the emission lines in Fig.\ref{f: forced}~B. Particularly noteworthy is the similarity of the locking observed for the LC's orbits with the phenomenon described in Ref.\,[S\onlinecite{Chafatinos2023}]. 



\subsection{Self-consistent phonons}
\begin{figure}
    \centering
    \includegraphics[width=0.4\linewidth]{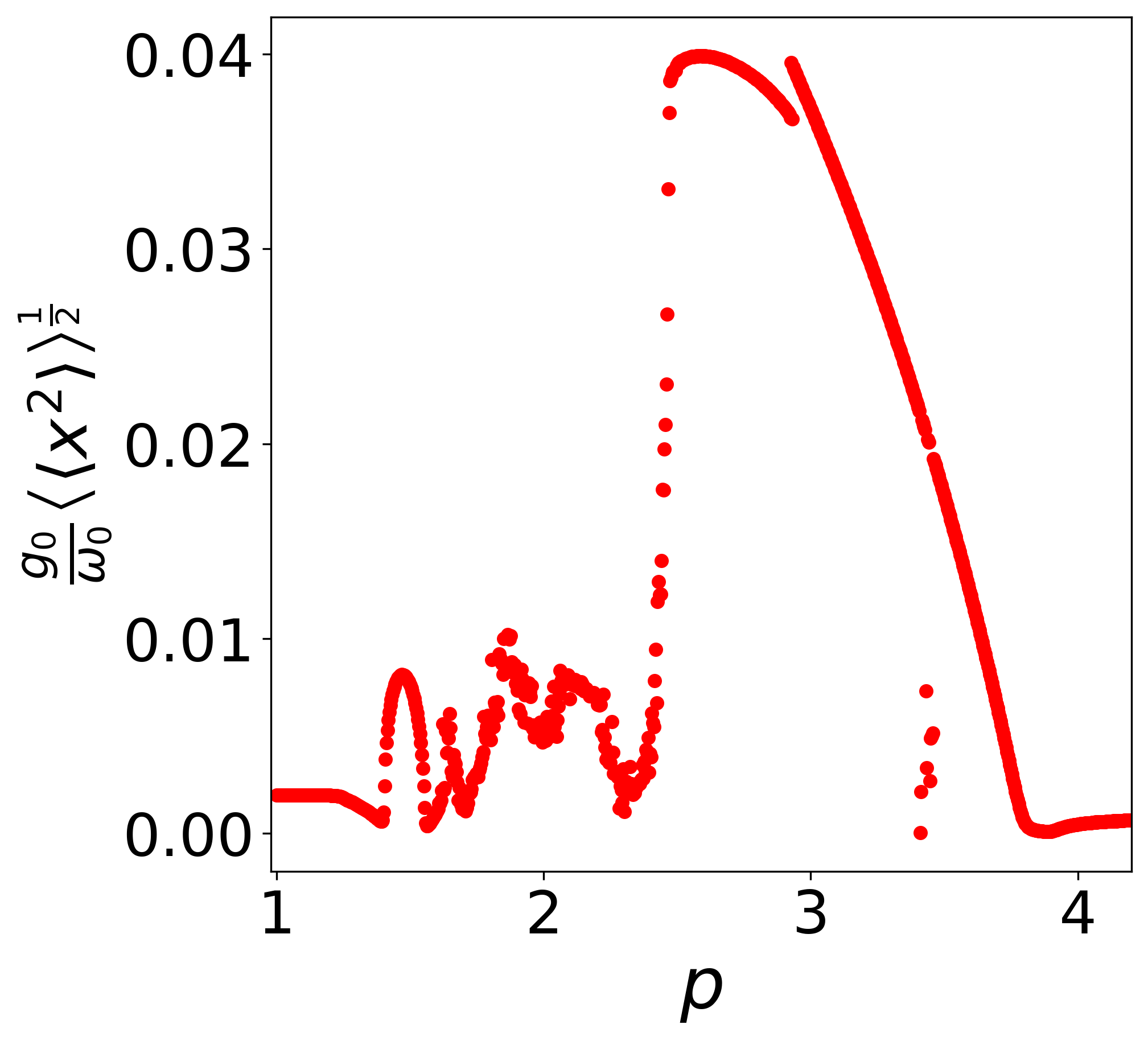}
    \caption{Time averaged phonon amplitude, $(g_0/\omega_0)\llangle x^2\rrangle^{\frac{1}{2}}$, as a function of the pump power $p$. }
    \label{fig:x_self}
\end{figure}

So far we have assumed that the phonons where present in the system. 
A self-consistent equation for the phonons can be derived using the equations of motion technique once the coupling Hamiltonian is defined \cite{Reynoso2022}. Following the same spirit as in the previous subsection, we assume that the coupling between polaritons and phonons comes from the modulation of the coupling constant $J$, which in turn can originate from a small difference of the optomechanical modulation of the X and Y modes. That is, $J\mapsto J_0 + \hbar g_0 \hat{x}$ where $\hat{x}=b+b^\dagger$ and $b$ ($b^\dagger$) the phonon annihilation (creation) operator. It follows immediately that, in the semiclassical limit when quantum correlations are ignored,  $x(t)=\langle \hat{x}\rangle$ satisfies the following equation
\begin{eqnarray}
\Ddot{x}&=&-\Gamma\Dot{x}-\omega_0^2x
+4\omega_0g_0\rho_0\mathrm{Re}(\tilde{\psi}_+^*\tilde{\psi}_-^{})\,,
\label{xn_effective_model}
\end{eqnarray}
that corresponds to a damped harmonic oscillator with a driving force provided by the polariton modes. It is important to keep in mind that this force is not externally given,  but rather obtained self-consistently by solving the gGPE \emph{simultaneously}.
While this approach is able to capture the locking phenomena described in the main text, it is an oversimplification of a very complex system, where different mechanism not fully considered here (highly excited states, phonon emission through relaxation mechanisms, etc) could in principle lead to some large fluctuation of the phonon field. 
We assumed to be this the case, and introduced in our equations a large initial condition.
The results presented in Fig. 3 of the main text were obtained using: $\Gamma/\omega_0=2.5\times10^{-4}$, $g_0/\omega_0=3.5\times10^{-5}$, $\rho_0\sim 2.5\times10^3$,  $x(0) \sim0.1\Omega_m/g_m$. Figure \ref{fig:x_self} shows the self-consistent value of $(g_0/\omega_0)\llangle x^2\rrangle^{\frac{1}{2}}$, that represents the modulation of $J$ in units of $\hbar\omega_0$, as a function of the pump power. Here $\llangle\dots\rrangle$ denotes time average. Note how the phonon amplitude is triggered in the region where the locking is strong and that the maximum amplitud of the modulation is of the order of $4$\% of $\hbar\omega_0$.

\section{Measurement of the linear polarization of the spinor}
\begin{figure}[t]
    \includegraphics[width=.7\linewidth]{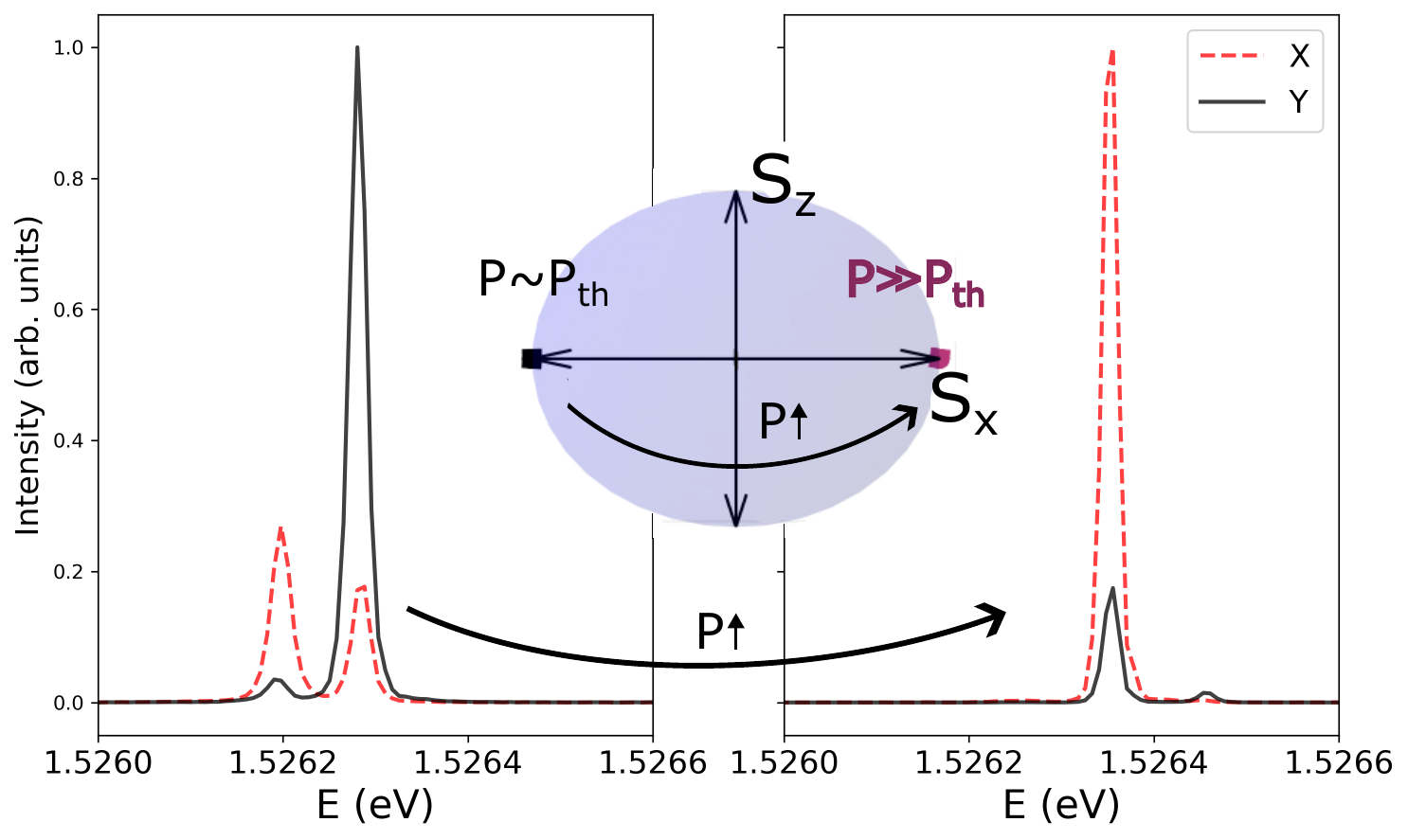}
    \caption{The X- and Y-linear polarized photoluminiscence for the ground state at $P=1.4P_{th}$ and $P=6.9P_{th}$ are presented in the left and right panel, respectively. The intensities have been normalized to the maximum value for each power. At low power (left panel) the predominant emission is Y-polarized, whereas at higher power (right panel) the principal emission is X-polarized. The central panel shows a simulation of the steady polariton states on a Poincare sphere showing two fixed points for negative (Y polarized) and positive (X polarized) axis of S$_x$ that correspond to power close to the power threshold ($P\sim P_{th}$) and significantly above threshold ($P\gg P_{th}$), respectively, and which aligns with the experimental results. Simulation parameters: $\gamma = 0.85,\ \gamma_R = 0.35,\ 2\gamma_d = 0.15,\ U\rho_0 = 0.25,\ U^R n_0 = 3.6,\ 2J = 0.45$.}
    \label{f: figExpSM2}
\end{figure}

In Fig.~\ref{f: figExpSM2}, we compare the vertical (Y, identified with continuous curves) and horizontal (X, dashed) linear polarization spectra corresponding to the ground state (GS) emission of a 2x2 $\mu$m$^2$ square trap for two different excitation powers, one close and above threshold ($P \sim P_\mathrm{th}$), the other much larger ($P \gg P_\mathrm{th}$) The spectra are normalized to the maximum emission at each power. The two emissions lines are associated with the spinor components of the polariton GS. The experimental linewidths are resolution-limited.

In the left panel, obtained for $P=1.4P_{th}$, the higher energy mode dominates the GS, resulting in Y-polarized emission. In contrast for the higher power at $P=6.9P_{th}$, presented in the right panel, the lower energy mode becomes the primary peak emitting predominantly X-polarized light. The inversion of the dominant emission in the spinor polariton states from X to Y with increasing power, is consistent with the model described in Eqs. \ref{eq: S-SM} and \ref{eq: nm}. This is consistent with our simulations, which are visualized 
on the central panel in Fig.~\ref{f: figExpSM2}, as two points on a Poincare sphere. 
For excitation powers close to the threshold, $P\sim P_\mathrm{th}$, the spinor stabilizes in a stable fixed point along $-S_x$, corresponding to linear polarized light along Y (black square marker). However, for significantly higher excitation powers ($P\gg P_{th}$), the spinor attains a stable fixed point along $S_x$, which corersponds to X-polarized light (violet square marker). A positive $\gamma_d$ was used to replicate the experiment. Conversely, for negative $\gamma_d$ the X-polarized state dominates at low powers, and as follows from our simulations the same polarization is stable at the higher powers.

\section{Additional experimental evidence of phonon-locked pseudospin dynamics and period doubling}

\subsection{Experiments on a resonator with $\lambda$ and $3\lambda$ confined phonons}

\begin{figure}[t]
    \includegraphics[width=.6\linewidth]{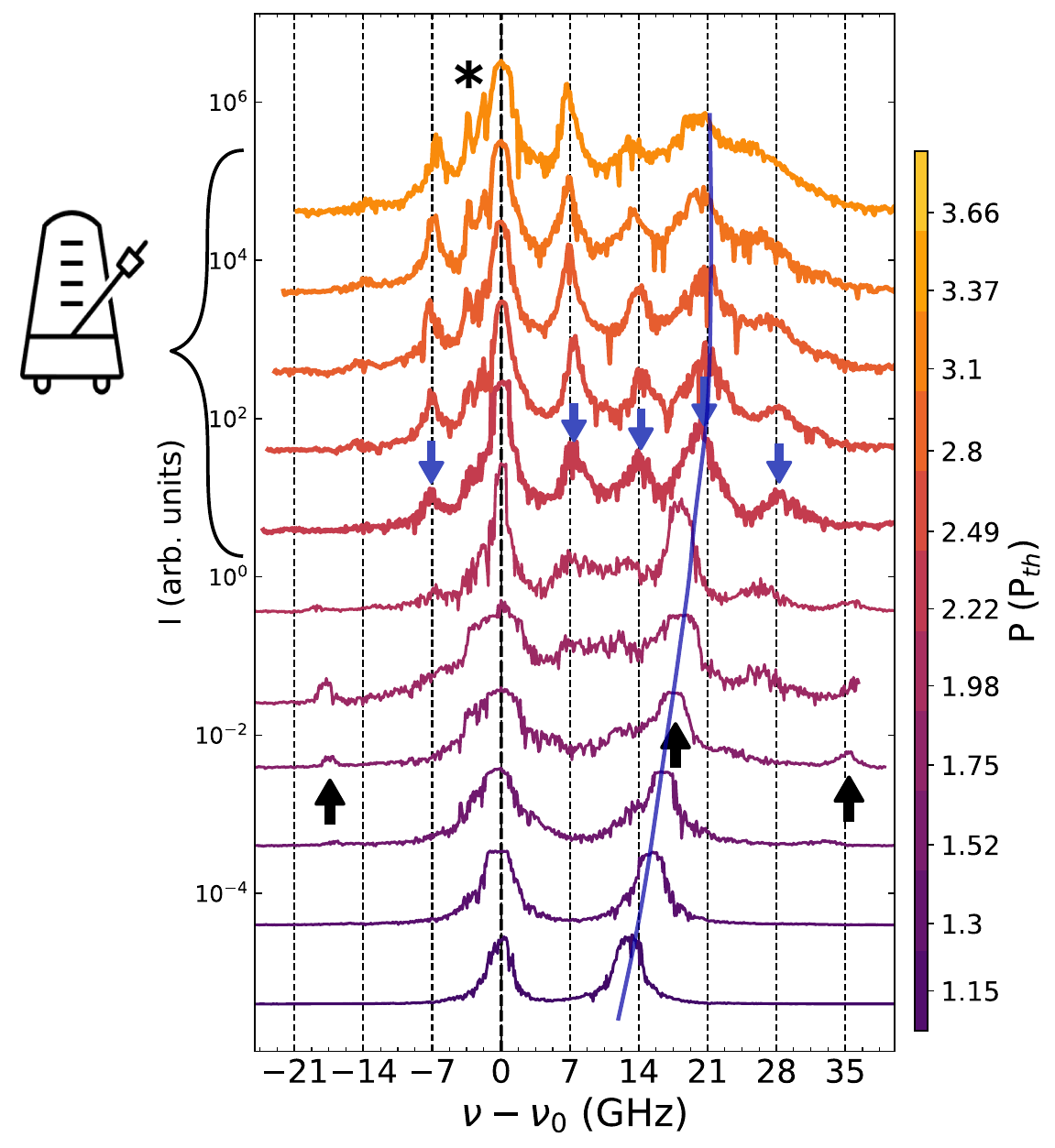}
    \caption{Normalized spectrum cascade of the ground spinor states for a 3x3$\mu$m$^2$ square trap. The two spinor modes initially exhibit a detuning of $\delta\nu\sim14$GHz, which increases with power. Black up arrows indicate a frequency comb resulting from Larmor precession. As the detuning approaches the cavity phonon frequency of $\nu_m\sim21$~ GHz, the spinor mode splitting locks and results in a new emission comb (marked by blue down arrows) at cavity phonon frequency $\nu_m\sim 7$ GHz corresponding to period tripling. The solid blue line is a guide to the eye indicating the locking to the mechanical clock. An asterisk denotes an artifact related to the tandem spectrometer. Note that for some spectra the main peaks were allowed to saturate the CCD to make the smaller features clearer. All spectra are normalized to the maximum intensity.}
    \label{f: figExpSM3}
\end{figure}

Here we present experiments on a different structure, sample B described in Section S1 of this supplementary material. This structure is designed so that polaritons can couple simultaneously with confined acoustic vibrations of frequencies $\nu^{\lambda}_m\sim 21$~GHz and $\nu^{3\lambda}_m\sim 7$~GHz.
In Fig.\,\ref{f: figExpSM3}, a sequence of photoluminescence spectra are presented with excitation power increasing from bottom to top. The spectral region corresponding to the ground state of polaritons confined within a 3x3 $\mu$m$^2$ square trap is displayed. All spectra that are shown correspond to powers that exceed the condensation threshold, $P>P_{th}$, resulting in sub-GHz linewidths for the modes. The emission was acquired with a tandem Fabry-Perot--Spectrometer set-up with resolution that allows for sub-GHz resolution.

The lower power (bottom-most) spectrum in Fig.~\ref{f: figExpSM3} is characterized by two emission lines corresponding to the spinor states which, at this power, are split by $\delta \nu \approx 14$~GHz. As theoretically discussed above, the origin of this splitting comes both from unintentional anisotropy of the trap (Josephson coupling $J$), and a possible spontaneous polarization of the reservoir leading through non-linearities to an additional effective magnetic field. As the excitation power is increased, the detuning between modes continuously augments and, quite notably, for the fourth spectrum from the bottom of the figure clear symmetric sidebands are developed (indicated with black up-arrows in the figure), with a separation that matches the mode detuning. The emergence of these sidebands is definitive proof of the transition to a dynamical state (time crystalline behavior) corresponding to a self-induced Larmor precession of the pseudospin. As the power is further increased, the line splitting, and hence the Larmor frequency, increases up to 21 GHz. When this situation is attained, quite notable changes occur. First, at around $P=2.49P_\mathrm{th}$, the detuning of the spinor states locks to the phonon frequency $\nu^{\lambda}_m\sim21$~GHz. This locking, which is evidence of the stimulation of a coherent mechanical oscillation induced by the polariton Larmor precession, is maintained up to the highest attained excitation power. Second, when this locking occurs it is accompanied by the emergence of a clear 7 GHz-frequency comb (indicated with blue downward arrows). That is, there is period tripling referred to the fundamental $\lambda$ phonon mode. Besides locking the polariton modes, presumably the self-induced $21$~GHz coherent phonons launch 7 GHz mechanical vibrations stabilizing a time crystal with a period of $1/\nu^{3\lambda}_m\sim 140$~ps.

Though somehow more complex due to the concurrence of two confined acoustic vibrations with frequencies differing by a multiple of 3, Fig.~\ref{f: figExpSM3} provides conclusive evidence first of the emergence of continuous time crystalline behavior characterized by Larmor precession of the spinor condensate, and later to its locking to a self-induced mechanical internal clock.

\subsection{Additional experiment and comparison with theory}

To highlight the universality of the reported time crystalline behavior, we present here experiments performed on still another structure, in this case with a more negative photon-exciton detuning (around 2~meV lower). The results have additional interest in that a quantitative modelling of the power-dependent spectra evidences the existence of Larmor orbits slightly different from those discussed in the main text.

\begin{figure}[t]
    \includegraphics[width=0.6\linewidth]{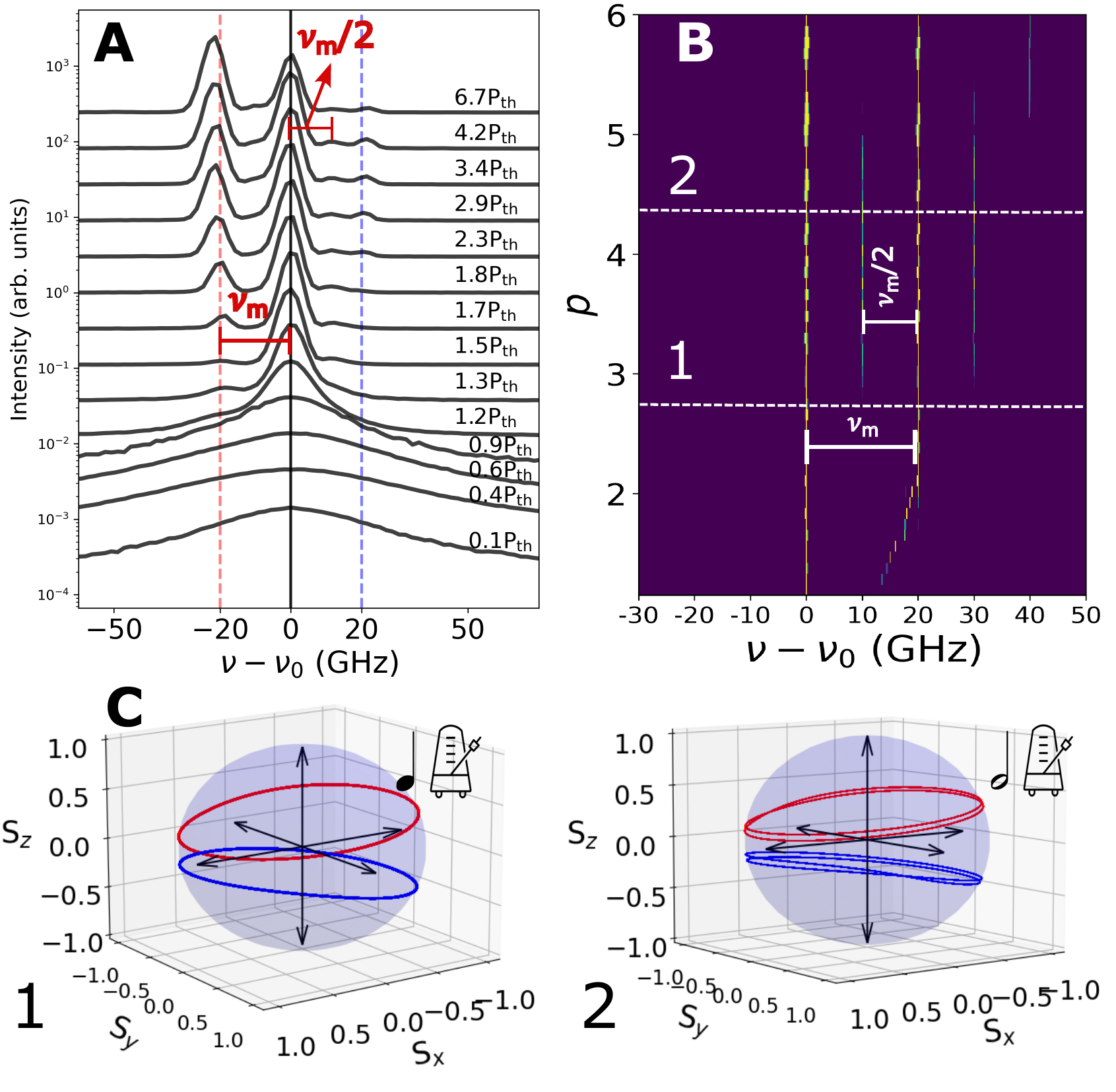}
    \caption{
\textbf{A} Cascade of photoluminescence normalized spectra corresponding to the spectral region around the ground state of a 2x2 $\mu$m$^2$ square trap. The curves span from $P<P_\mathrm{th}$ to $P \gg P_\mathrm{th}$. \textbf{B} Calculated spectra for varying excitation power $p$. The trivial blueshift has been subtracted from both experimental and theoretical spectra. Above $P_\mathrm{th}$ two main peaks are clearly resolved split by the phonon frequency $\nu_{ph} \sim 20$GHz, and sidebands appear at half of the phonon frequency $\nu_{ph}/2 \sim 10$GHz. Simulation parameters: $\gamma = 0.25,\ \gamma_R = 0.23,\ \gamma_d = -0.05,\ U\rho_0 = 0.5,\ U^R n_0 = 7.2,\ J_0 = 0.14$, and $J_p = 0.015$. \textbf{C} Normalized Poincar\'e spheres for different orbits corresponding to the powers marked with white horizontal dashed lines in \textbf{B}. Two possible degenerate limit cycles for each spectrum, approached depending on the initial conditions, are drawn in red for the upper hemisphere and blue for the lower hemisphere.}
    \label{f: figExpSM5}
\end{figure}

Figure~\ref{f: figExpSM5}\,A presents experimental spectra obtained again around the ground state's spectral region of a 2x2 $\mu$m$^2$ square trap. The excitation power increases from bottom to top, as indicated in the figure. For $P < P_{\mathrm{th}}$(the first four spectra), a single broad emission line is observed. Subsequently, with increasing power this line undergoes a sudden narrowing (limited by the spectrometer's resolution), signaling the onset of condensation. When this occurs two lines become clearly discernible, separated at $\sim$~20 GHz (the phonon frequency). This same splitting is maintained up to the highest attained excitation power (locking to the mechanical clock). Quite notably, for $P \gtrsim 2P_{th}$), additional emission lines appear spaced at half the mechanical frequency $\sim 10$~GHz (period doubling with continuous excitation). 

The observed behavior is similar but not exactly equal to that of a different  2x2 $\mu$m$^2$ square trap presented in Fig.\,2A of the main text (being the one in the main text more excitonic). Here in contrast there are two clear and distinguishable more intense peaks, which maintain a constant separation equal to the phonon frequency ($\sim$20 GHz) almost throughout all the power range. To understand these subtle differences, we modeled the system using Eqs. \ref{eq: S-SM}, \ref{eq: BE-SM}, and \ref{eq: nm} with a forced optomechanical modulation ($J(t) = J_0 + J_{ph} \cos{(t)}$). The parameters were adjusted to better describe the observed behavior, mainly decreasing the cavity dissipation to account for the more photonic character (the used parameters are reported in the caption to Fig.~\ref{f: figExpSM5}\,A). The power dependence of the calculated spectra are presented in Fig.~\ref{f: figExpSM5}\,B. The resemblance with the experiments is noteworthy. Throughout the entire range of applied powers $p$, two main peaks are present. As the excitation power increases ($p\sim 1-2.3$), rapidly the separation, and consequently the dynamic frequency, grows until it matches the phonon's frequency. Once this occurs, the Larmor frequency synchronizes with the mechanics. The calculated orbits on the normalized Poincare sphere are shown in Fig.~\ref{f: figExpSM5}\,C1 corresponding to this situation ($p = 2.8$). There are two possible degenerate solutions that depend on the initial conditions. The two are characterized with distinct polarizations, distinguished with different colors, red for the one on the higher hemisphere and blue for the one in the lower hemisphere.

Continuing with the theoretical simulation, around $p\sim 3-5.2$ small sidebands emerge accompanying the main peaks, separated by half the phonon frequency. Notably, similar to the experimental spectra these sidebands persist for the whole power range, and appear predominantly in between and on the higher frequency side of the main peaks. This different kind of behavior has also implications on the orbits of the existing limit cycles, as shown in Fig.~\ref{f: figExpSM5}\,C2. There is again a period doubling as described in the main text, the orbits close after two turns, but with a subtle twist. Here two solutions persist in which the two turns are very close one to the other and to the Larmor orbit existent before period doubling (they reside in the same hemisphere). In contrast, the double orbits described in the main text correspond to the merging of the two possible Larmor orbits, the spinor visits the two hemispheres on a period, and thus this implies a stronger modulation of the reservoir and polariton polarization.

\subsection{High power instability}

We present here some additional results that further confirm the observation of time crystalline behavior of a polariton condensate in a trap, locking and period doubling, but that add some interesting peculiarities.  These were again observed for the 2x2 $\mu$m$^2$ square trap with larger excitonic component described in the main text, but exploring higher excitation powers. In addition, while for the experiments in the main text the excitation laser spot was slightly displaced to the side of the traps to reduce the reservoir induced noise, in the experiments described in this section it was positioned closer to the center of the trap.

\begin{figure}[t]
    \includegraphics[width=.7\linewidth]{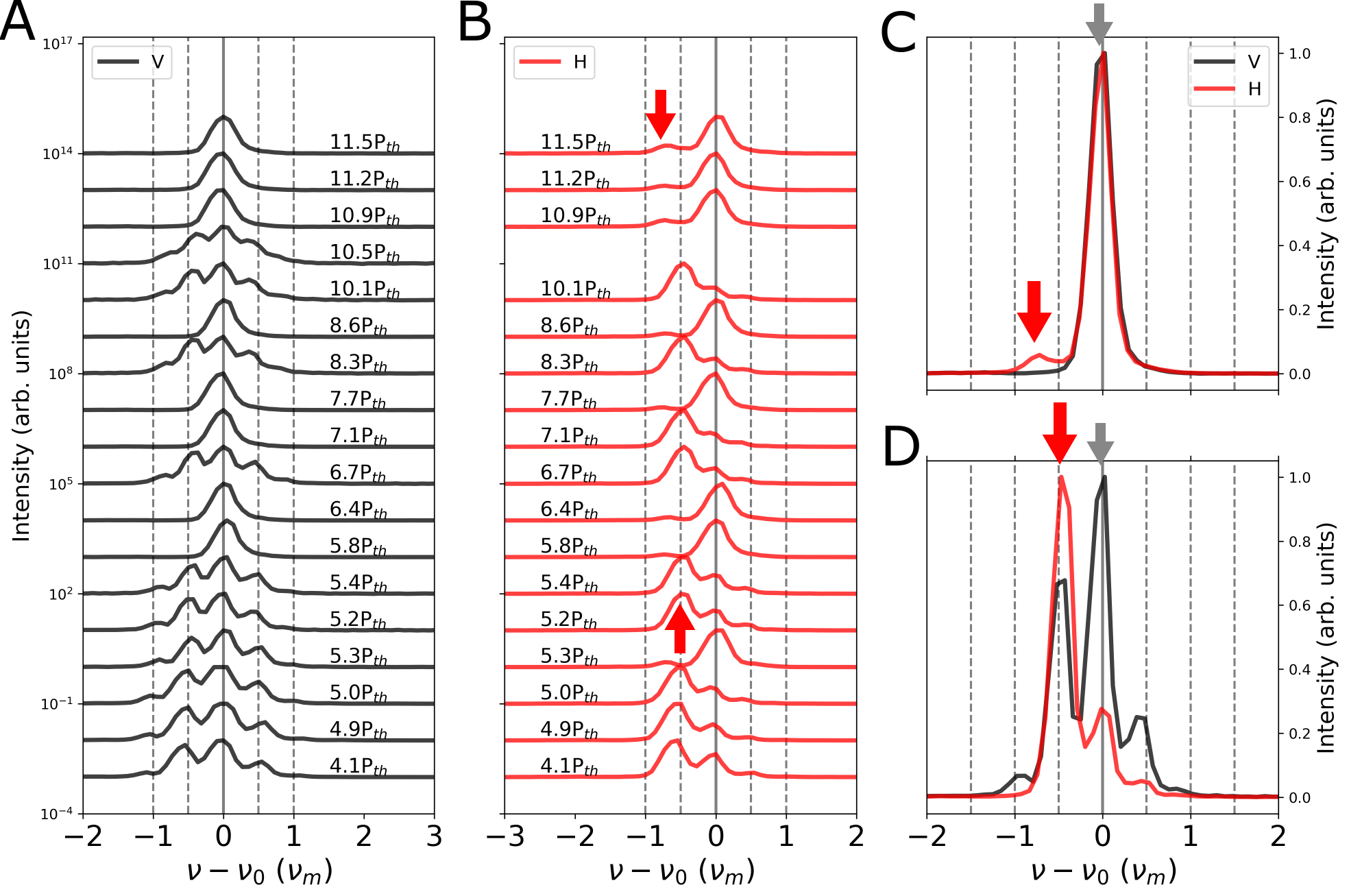}
    \caption{Normalized polariton spectra around the spectral region corresponding to the ground state of a 2x2 $\mu$m$^2$ square trap, for the same structure described in Fig.~1G of the main text. The shown cascade corresponds to increasing excitation power from bottom to top, indicated in each curve. The emission was filtered with a linear polarizer, vertical in \textbf{A} and horizontal in \textbf{B}. A metastability is observed with the spectra jumping between two situations which are illustrated in \textbf{C} and \textbf{D}. The main emission lines are indicated with red and gray arrows (see text for details).}
    \label{f: figEXPSM4}
\end{figure}

Figures~\ref{f: figEXPSM4}A and B present the photoluminescence experiments reaching $P = 11.5 P_{\mathrm{th}}$. The two figures correspond to two different filtered linear polarizations, vertical and horizontal for A and B, respectively. The first notable result, is that there is an instability for the higher powers, with the spectra jumping between two different shapes. Notably, this instability occurs in pairs, each kind of spectra for one polarization accompanied by the same kind of spectra for the other polarization. The two observed situations are illustrated in Figs.~\ref{f: figEXPSM4}\,C and D. In C only one main peak is observed (the higher frequency one) for the two polarizations, suggesting the existence of a stable fixed point with polarization rotated towards a diagonal of the square trap. Interestingly, in D there are two main peaks characteristic of Larmor precessing orbits, and this is additionally evidenced by the emergence of clear sidebands separated by half the phonon frequency, signaling period doubling. 

What is more striking in these spectra is that when sidebands are observed, the splitting of the two main peaks is locked to the phonon frequency $\sim 20$~GHz, while the metastable state without sidebands corresponds to the splitting of the spinor states reduced to $\sim 15$~GHz. We interpret this result as compelling evidence that the mechanics stabilizes the time crystalline behavior, and particularly the observation of period doubling in the absence of an external time-dependent driving.

\section{Experimental evidence of dissipative coupling}


\begin{figure}[t]
    \includegraphics[width=.4\linewidth]{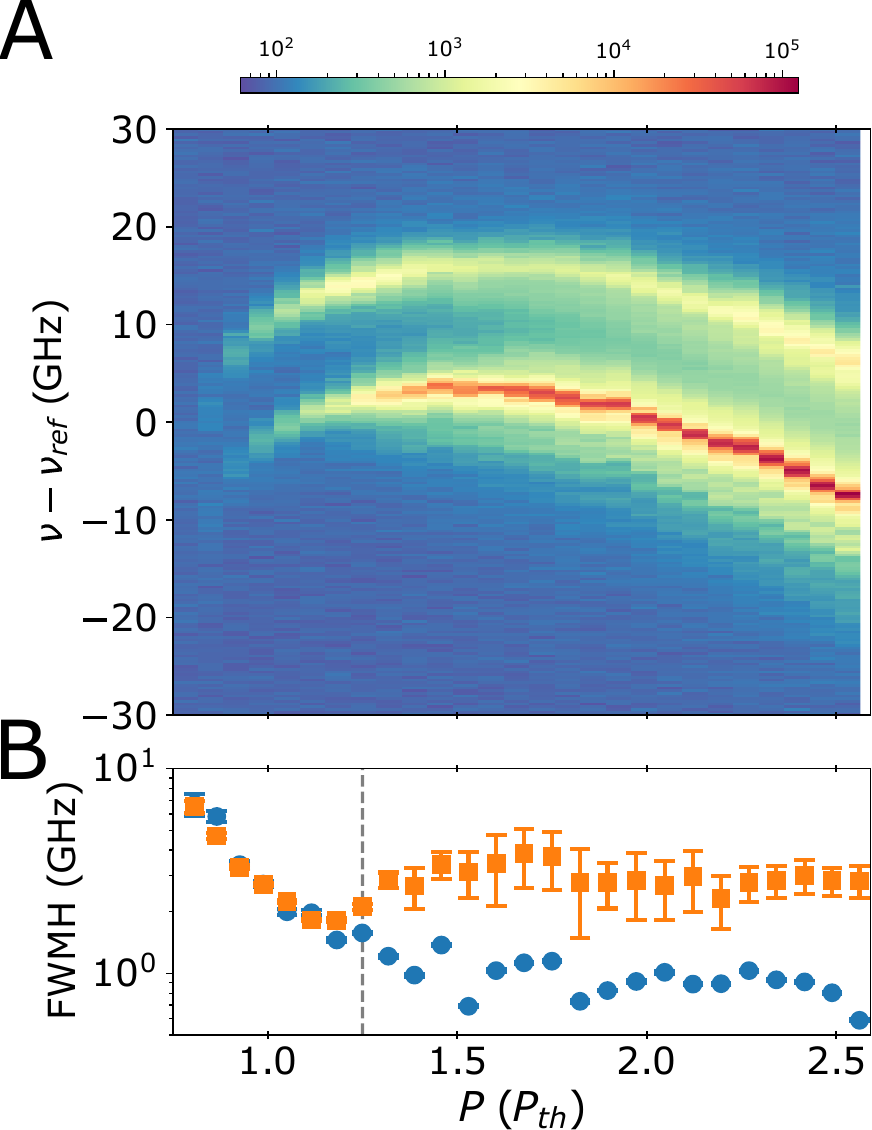}
    \caption{\textbf{A}: Photoluminescence (PL) spectra of a 4x4 $\mu$m$^2$ square trap around the ground state spectral region,  shown as a function of excitation power. Two modes are identified. \textbf{B}: Full width at half maximum as a function of excitation power. The upper and lower modes are represented by triangle and circle markers, respectively. Notably, at 1.5P$_{th}$, the upper mode suddenly exhibits a broader linewidth compared to the lower mode. This indicates the emergence of effective dissipative coupling.}
    \label{f: figExpSM1}
\end{figure}

The upper panel of Fig.\,\ref{f: figExpSM1} illustrates the emission spectra of the spinor's ground state for a 4x4 $\mu^2$ trap as a function of excitation power. Two distinct emission lines are identified, and attributed to the split spinor levels. These spectra correspond to sample B, and the levels appear separated close to 14~GHz, that is twice the frequency of the $\nu^{3\lambda}_m\sim7$~GHz acoustic phonon mode.

The lower panel of Fig. \ref{f: figExpSM1} presents the corresponding linewidths, given by the full width at half maximum. Circles (squares) correspond to the lower (upper) energy mode. At the lower powers, just above threshold signaled by the rapid narrowing with applied excitation power, both lines share the same width. However, a point is reached where the width of the two lines abruptly splits. The upper mode suddenly changes its behavior displaying an additional strong broadening. The linewidth difference saturates at higher powers.

The sudden difference in linewidths can be linked to the process of spontaneous symmetry breaking, where the time-averaged reservoir magnetization $\left|\left<m\right>\right|$ acquires a finite value, as demonstrated in the lowest panel of Fig.\,\ref{f: figSM3}. The difference in linewidth is attributed to the dissipative coupling terms in the synthetic crystal field $\bm{E}$ (Eq.\,\eqref{eq: BE-SM}). The $-2\gamma_d\xv$ component represents a constant dissipative coupling due to the disparity in dissipation between the $X$ and $Y$ linear polarization modes. Conversely, the $\gamma m \zv$ component varies depending on the solutions of the system. When $m$ takes a finite constant component, it results in a difference in gain between the + and - modes.

\section{A polariton condensate discrete time crystal induced by a pulsed by ps-laser}

As can be expected from the theoretical considerations in this supplementary material, and in the main text, it is expected that the system exhibits period-doubling behavior when subject to an external mechanical driving force. This is in contrast with the self-induced dynamics described up to here. To test this prediction, we generate coherent phonons using a ps-acoustics technique based on the impulsive excitation of vibrations with an ultrafast pulsed laser.

\begin{figure}[ht]
    \centering 
    \includegraphics[trim = 0mm 0mm 0mm 0mm,clip=true, keepaspectratio=true, width=0.6\columnwidth,angle=0]{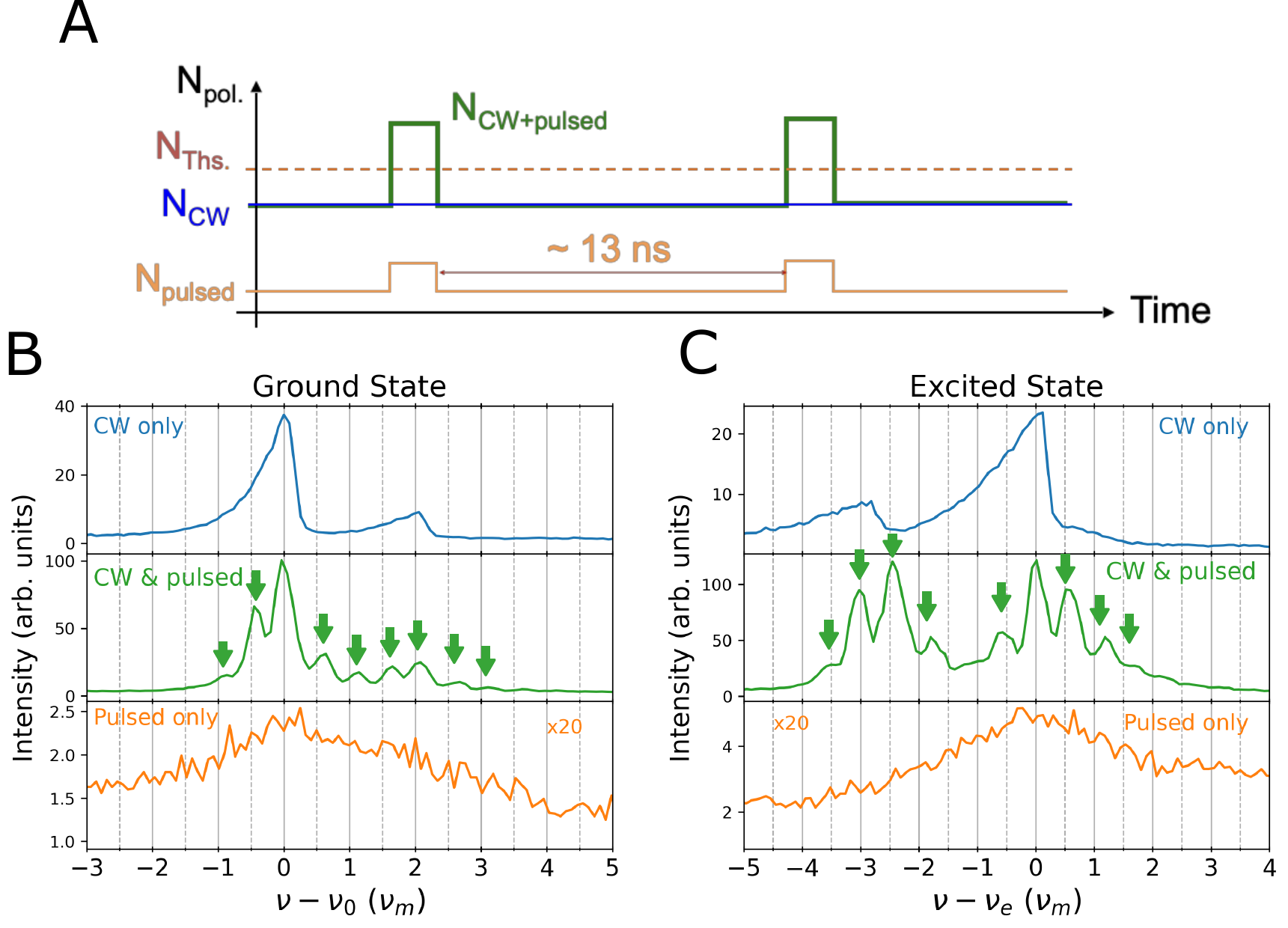}
    \caption{ \textbf{A} Scheme of the continuous wave (cw) plus pulsed laser excitation protocol. A cw laser is used to populate the polariton ground and first excited state of a $4 \times 4~\mu$m$^2$ square trap, with excitation power just {\em below} condensation threshold (P$_{Th}$). A second weaker ps-laser provides the power which added to that of the cw laser surpasses P$_{Th}$ with $\sim 80$MHz repetition rate (pulses impinging every $\sim 13$ns). Such pulsed ps-excitation is known to launch coherent phonons which, for this cavity structure, were designed to occur at $\omega_m/2\pi \sim 7$ GHz. \textbf{B}(\textbf{C}) Time-integrated spectra collected with the pulsed laser only (bottom), cw laser only (top), or both (middle spectrum) for the polariton ground state [first excited state]. Note the emergence of sidebands at integer multiples of $\nu_m/2 \sim 3.5$ GHz (highlighted with down arrows), a period doubling signaling the emergence of a discrete time crystalline state.}
    \label{Fig4}
\end{figure} 

The experimental protocol outlined in Fig.~\ref{Fig4}\,A unfolds as follows: (i) initially, a non-resonant continuous-wave laser excitation, with power deliberately set slightly below P$_{th}$, places the system on the brink of polariton condensation; (ii) subsequently, a pulsed laser emitting $\sim 1$~ps pulses every $\sim 13$~ns is introduced. This laser not only supplies the additional power necessary for condensation but, more crucially, impulsively generates coherent mechanical vibrations at the frequency $\omega_{ph}$. The repetition rate of this second laser is chosen to ensure that the impulsively generated coherent vibrations relax before the arrival of the subsequent pulse.

This experiment were conducted on an individual square trap measuring $4 \times 4\mu$m$^2$. Due to its larger dimensions, this trap confines two polariton modes (ground and excited) separated by $\sim 2$~meV ($\sim 480$GHz$ \gg \omega_{ph}/2\pi$). Figure~\ref{Fig4}\,B and C display the resulting time-integrated spectra corresponding to the ground and excited polariton states of the trap, respectively, under three conditions: only continuous-wave excitation (top), only pulsed excitation (bottom), and both excitations applied simultaneously (middle). Importantly, several narrow sidebands separated by $\omega_{ph}/2$ are distinctly observable, influencing both the ground and excited states when both lasers are present. This unequivocally establishes the existence of a discrete time crystal in the context of a Floquet-like mechanical driving of the polariton condensate.

\end{document}